\documentclass{sig-alternate-10pt}
\usepackage{url}
\usepackage{amsmath}
\usepackage{amssymb}
\usepackage{bm}
\usepackage{stmaryrd}
\usepackage{color}
\usepackage{colortbl}
\usepackage{epsf}
\usepackage{supertabular}
\usepackage{hhline}
\usepackage{multicol}
\usepackage{times}

\newcommand{\ignore}[1]{}

\newcommand{\rred}{}
\newcommand{\re}{}
\newcommand{\bblue}{}

\newcommand{\blue}{}

\ignore{
\newtheorem{proposition}{Proposition}
\newtheorem{lemma}{Lemma}
\newtheorem{corollary}{Corollary}
\newtheorem{theorem}{Theorem}
\newtheorem{definition}{Definition}

\newtheorem{example}{Example}
\newtheorem{remark}{Remark}   }

\DeclareMathAlphabet{\mathpzc}{OT1}{pzc}{m}{it}

\def\be{\begin{equation}}
\def\ee{\end{equation}}
\def\bea{\begin{eqnarray}}
\def\eea{\end{eqnarray}}
\def\nn{\nonumber}

\def\w{\wedge}
\def\ra{\rightarrow}

\def\P{\Phi}

\def\bc{\bigcap}
\def\S{\Sigma}
\def\nin{\not\in}

\def\ex{\exists}

\def\sc{\scriptsize}

\def\ba{\begin{align}}
\def\ea{\end{align}}
\def\bes{\begin{split}}
\def\es{\end{split}}

\def\ems{\emptyset}

\newcommand{\red}[1]{{\em \bf\textcolor[rgb]{1.00,0.00,0.00}{#1}}}
\ignore{
\newcommand{\red}[1]{{\em \bf\textcolor[rgb]{1.00,0.00,0.00}{#1}}}
\newcommand {\blue}[1]{\textcolor[rgb]{0.00,0.00,1.00}{#1}}

}

\newcommand{\comlb}[1]{{\vspace{2mm}\noindent \bf \red{COMM(LEO):}}~ #1 \hfill {\bf
    END.}\\}
\newcommand{\comj}[1]{{\vspace{2mm}\noindent \bf \red{COMM(JAF):}}~ #1 \hfill {\bf
    END.}\\}

    \newcounter{theorem-counter}
\newcounter{corollary-counter}
\newcounter{lemma-counter}
\newcounter{definition-counter}
\newcounter{example-counter}
\newcounter{proposition-counter}
\newcounter{remark-counter}

\newenvironment{theorem}%
{\vskip \abovedisplayskip \refstepcounter{theorem-counter}%
\noindent {\bf Theorem \arabic{theorem-counter}.}}%

\newenvironment{corollary}%
{\vskip \abovedisplayskip \refstepcounter{corollary-counter}%
\noindent {\bf Corollary \arabic{corollary-counter}.}}%
{\boxtheorem}

\newenvironment{lemma}%
{\vskip \abovedisplayskip \refstepcounter{lemma-counter}%
\noindent {\bf Lemma \arabic{lemma-counter}.}}%

\newenvironment{definition}%
{\vskip \abovedisplayskip \refstepcounter{definition-counter}%
\noindent {\bf Definition \arabic{definition-counter}.}}%
{\newline}

\newenvironment{example}%
{\vskip \abovedisplayskip \refstepcounter{example-counter}%
\noindent {\bf Example \arabic{example-counter}.}}%

\newenvironment{proposition}%
{\vskip \abovedisplayskip \refstepcounter{proposition-counter}%
\noindent {\bf Proposition \arabic{proposition-counter}.}}%

{\vskip \abovedisplayskip \refstepcounter{remark-counter}%
\noindent {\bf Remark \arabic{remark-counter}.}}%

\newcommand{\defproof}[2]{{\noindent\bf Proof of #1:\
}#2 \boxtheorem}

\newcommand{\boxtheorem}{\hfill $\Box$ \vspace{2mm}}
\newcommand{\nit}[1]{{\it #1}}

\newcommand{\mc}[1]{\mathcal{ #1}}

\begin{document}
\thispagestyle{empty}

\title{{\bf \Large Tractable vs. Intractable Cases of Matching Dependencies for Query Answering under Entity Resolution}\thanks{An extended abstract containing
in preliminary form some of results in this paper was presented at
the Alberto Mendelzon WS on Foundations of Data Management, 2013.}}

\author{
{\bf \large Leopoldo Bertossi}           \hspace{2.3cm}         {\bf \large Jaffer Gardezi}         \\
Carleton University, SCS \hspace{0.9cm} University of Ottawa, SITE.\\
Ottawa, Canada              \hspace{2.2cm}               Ottawa, Canada\ignore{\\
jgard082@uottawa.ca \hspace{1.7cm} bertossi@scs.carleton.ca}
}

\maketitle
\begin{abstract}
Matching Dependencies (MDs) are a relatively recent proposal for
declarative entity resolution. They are rules that
specify, on the basis of similarities satisfied by values in
a database, what values should be considered duplicates, and have to be matched.
On the basis of a  chase-like procedure for MD enforcement, we can obtain
clean (duplicate-free) instances; actually possibly several of them. The resolved answers to queries
are those that are invariant under the resulting class of resolved instances.
Previous work identified certain classes of queries and sets of MDs for which resolved query answering is tractable. Special emphasis
was placed on cyclic sets of MDs.
In this work we further  investigate
the complexity of this problem, identifying intractable cases, and exploring the frontier between tractability and intractability. We concentrate
mostly on acyclic sets of MDs.  For a special case we obtain a dichotomy
result relative to {\it NP}-hardness.
\end{abstract}

\vspace{2mm}\noindent{\bf Keywords:} \ data cleaning, entity resolution, matching dependencies, query answering, data complexity

\section{Introduction}

A database may contain several representations of the same external entity. In this sense it
contains ``duplicates", which is in general considered to be undesirable; and
the database has to be cleaned. More precisely,
the problem of {\em duplicate- or entity-resolution} (ER) is about (a) detecting duplicates, and (b) merging
duplicate representations into single representations.
This is a classic and complex problem in data management, and in data cleaning in particular \cite{naumannACMCS,elmargamid,BenjellounGMSWW09}.
In this work we concentrate on the merging part of the problem, in a relational context.

A generic way to approach the problem consists in specifying what attribute values have to be
matched (made identical) under what conditions. A declarative language with a precise semantics could be used for this purpose.
In this direction, \re{matching dependencies} (MDs) have been recently introduced \cite{Fan08,Fan09}.
They represent rules for
resolving pairs of duplicate representations (considering two tuples at a time). Actually, when certain similarity relationships between
attribute values hold, an MD indicates what attribute values
have to be made the same (matched).

\begin{example} \  The similarities of phone and address
indicate that the tuples refer to the same person, and the
names should be matched. Here, {\small 723-9583 $\approx$ (750) 723-9583}   and   {\small \bblue{10-43 Oak St.} $\blue{\approx}$ \bblue{43 Oak St. Ap. 10}}.

\begin{center}
{\small \begin{tabular}{c|c|c|c|}\hline
$\blue{\nit{People}}$ & Name & Phone & Address \\ \hline
&\rred{John Smith} & \bblue{723-9583} & \bblue{10-43 Oak St.} \\
&\rred{J. Smith} & \bblue{(750) 723-9583} & \bblue{43 Oak St. Ap. 10}  \\ \hhline{~---}
\end{tabular}  }
\end{center}

The following MD captures this resolution policy: \ (with $P$ standing for predicate $\nit{People}$)
\begin{eqnarray*}
P[\nit{Phone}]\approx P[\nit{Phone}] \! &\w&\!P[\nit{Address}]\approx P[\nit{Address}]  \ra\\
&& ~~~~~~~~P[\nit{Name}]\doteq P[\nit{Name}].
 \end{eqnarray*}
This MD involves only one database predicate, but in general,
an MD may involve two different relations. We can also have several MDs on the database schema. \boxtheorem
\end{example}
The framework for MD-based entity resolution used in this paper was introduced in \cite{front12},
where a precise semantics for MDs involving a chase procedure for
cleaning the database instance was introduced. This semantics made precise the
rather intuitive semantics for MDs originally introduced in \cite{Fan09}.

Also in \cite{front12}, the
problem of {\em resolved query answering} was introduced. For a fixed set of MDs, and a fixed query,
this is the problem
of deciding, given an  unresolved database instance, and a candidate query answer $\bar{a}$,
whether $\bar{a}$ is an answer to the query under all admissible ways of resolving the duplicates
as dictated by the MDs. It was shown that
this problem is generally intractable by giving an \nit{NP}-hard case of
the problem involving a pair of MDs. By identifying the elements of this
set of MDs that lead to
intractability, tractability of resolved query answering was obtained for
other pairs of MDs.

The resolved query answering problem was studied further in \cite{sum12,datalog12}.
Specifically, a class of tractable cases of the problem was identified \cite{sum12}, for
which a method for retrieving the resolved answers based on
query rewriting into stratified Datalog with aggregation was developed
\cite{datalog12}.

In those tractable cases, we find conjunctive queries
with certain restrictions on joins, and sets of MDs that depend cyclically on each other,
in the sense that modifications produced by one MD may affect the application of the
next MD in the enforcement cycle. These are  the (cyclic) {\em HSC sets} identified in \cite{sum12}. It was shown that, in general, cyclic dependencies on MDs make the problem
tractable, because the requirement of chase termination implies a relatively simple
structure for the clean database instances \cite{sum12}.

In this work we concentrate on acyclic sets of MDs.
This completely changes the picture wrt. previous work. As just mentioned, for HSC sets, tractability of resolved query answering holds \cite{sum12}. This is
the case, for example, for the cyclic $M = \{R[A] \approx R[A] \ra R[B] \doteq R[B], \ R[B] \approx R[B] \ra R[A] \doteq R[A]\}$. However, as we will see later on, for the following acyclic, somehow syntactically similar
example, $M' = \{R[A] \approx R[A] \ra R[B] \doteq R[B], \ R[B] \approx R[B] \ra R[C] \doteq R[C]\}$, resolved query answering can be intractable.
This example, and our general results, show
that, possibly contrary to intuition, the presence of cycles in
sets of MDs tends to make resolved query answering easier.

In this work, we further explore the complexity of resolved query answering.
Rather than considering isolated intractable cases as in previous work,
here we take a more systematic approach. We develop a set of
syntactic criteria on sets of two MDs that, when satisfied by a given pair of
MDs, implies intractability of the resolved query answer problem.

We also show,
under an additional assumption about the nature of the similarity operator,
that resolved query answering is tractable for sets of MDs not satisfying these
criteria, leading to a dichotomy result. We extend these results also
considering tractability/intractability of sets of more than two MDs.

All these results apply to acyclic sets of MDs,
and thus are complementary to those of \cite{sum12,datalog12}, providing a broader
view of the complexity landscape of query answering under matching dependencies.

Summarizing, in this paper, we undertake a systematic investigation of the data complexity of the problems of deciding and computing
resolved answers to conjunctive queries under MDs. This complexity analysis sheds some light on the intrinsic computational limitations of retrieving, from a database
with unresolved duplicates,
the information that is invariant under the entity resolution processes as captured by
MDs. The main contributions of this paper are
as follows:
\begin{enumerate}
\item We identify a class of conjunctive queries that are relevant for the investigation of
tractability vs. intractability of resolved query answering. Intuitively,
these \linebreak queries return data that can be modified by
application of the MDs. We call them {\em changeable attribute
queries}.

\item Having investigated in \cite{datalog12,sum12} cases of cyclic sets of MDs, we complement these results
by studying the complexity of resolved query answering for sets of MDs that do not have cycles.

\ignore{a dependency graph (MD graph) was associated with a set of MDs, and
it was shown that the resolved answer problem can be tractable for
a broad class of conjunctive queries when there are cycles in this
graph. Sets of MDs with such cyclic MD
graphs are called {\em Hit-Simple-Cyclic} (HSC) sets of MDs.
 In Section \ref{sec:hardacyclic}, we complement these results
by studying the complexity of sets that do not have cycles in their MD graphs.}

\item For certain sets of two MDs that satisfy a syntactic condition, we establish an intractability result, proving that deciding
resolved answers to changeable attribute queries is
${\it N\!\!P}$-hard in data.

\item For similarity relations that are transitive (a rare case), we establish that the conditions for hardness mentioned in the previous item,
lead to a dichotomy result: pairs of MDs that satisfy them are always {\em hard}, otherwise they are always {\em easy} (for resolved query answering).
This shows, in particular, that the result mentioned in item 3.
cannot be extended to a wider class of MDs for arbitrary similarity relations.

We also prove that the dichotomy result does not hold when the hypothesis on similarity is not satisfied.

\item Relying on the results for pairs of MDs, we consider acyclic sets of MDs of arbitrary size.
In particular, we prove intractability
of the resolved query answering problem for certain
acyclic sets of MDs that have the syntactic property of {\em non-inclusiveness}.
\end{enumerate}

The structure of the paper is as follows. Section \ref{sec:prel} introduces
notation and terminology used in the paper, and reviews necessary results
from previous work. Section \ref{sec:new} identifies classes of MDs, queries and assumptions that are relevant for this research.
Sections \ref{sec:linpairs} and \ref{sec:dich}
investigate the complexity
of the problem of computing resolved answers for sets of
two MDs. Section \ref{sec:hardacyclic} extends those results to
sets of MDs of arbitrary size.
 In Section \ref{sec:disc} we summarize results, including a table of
known complexity results (obtained in this and previous work). We also draw some final conclusions, and we point to open problems.
Full proofs of our results can be found in the appendix.\footnote{An extended abstract containing
in preliminary form some of results in this paper is \cite{amw13}.}

\section{Preliminaries}\label{sec:prel}

In this work we consider relational database schemas and instances. Schemas are usually denoted
with $\mc{S}$, and contain relational predicates. Instances are usually denoted with $D$.
Matching dependencies (MDs) are symbolic rules of the form: \vspace{-2mm}
\begin{equation}\label{eq:md}
\bigwedge_{i , j }R[A_i] \approx_{ij} S[B_j]  \ \ra \ \bigwedge_{k, l}R[A_k]\doteq S[B_l],
\end{equation}
where $R, S$ are relational predicates in $\mc{S}$, and the $A_i, ...$ are attributes for them. The LHS captures similarity conditions on a pair of tuples belonging to the extensions of $\blue{R}$ and $\blue{S}$
in an instance $D$. We  abbreviate this formula as: \ $
R[{\bar A}] \approx S[{\bar B}] \ \ra \ R[\bar C]\doteq S[\bar E]
$.

The similarity predicates (or operators) \bblue{$\approx$} (there may be more than one in an MD depending on the attributes involved) are domain-dependent and treated as built-ins. Different attribute domains may have different similarity
predicates. We assume  they are symmetric and reflexive. Transitivity is not assumed  (and in many applications it may not hold).

MDs have a {\em dynamic interpretation} requiring that those values on the RHS
should be updated to some (unspecified) common value. Those attributes on a RHS of an MD are called \bblue{\em changeable attributes}.
MDs are expected to be ``applied" iteratively until duplicates are solved.

In order to
keep track of the changes and comparing tuples and instances, we use global
tuple identifiers, a non-changeable surrogate key for each database predicate that
has changeable attributes. The auxiliary, extra attribute (when shown) appears as the first
attribute in a relation, e.g. \bblue{$t$} is the identifier in \bblue{$R(t,\bar x)$}.
A \bblue{\em position} is a  pair \bblue{$(t,A)$}
with \bblue{$t$} a tuple id, and $A$ an attribute (of the relation where $t$ is an id).
The \blue{\em position's value}, \bblue{$t[A]$}, is  the value for $A$ in tuple
(with id) \bblue{$t$}.

\subsection{MD semantics}\label{sec:sem}

A semantics for MDs acting on database instances was
proposed in \cite{front12}.
It is based on a {\em chase procedure} that is iteratively applied to the original instance \bblue{$D$}.
A \bblue{\em resolved instance}
\bblue{$D'$} is obtained from a finitely terminating
sequence
of database instances, say
\begin{equation}
D =: D_0 \ \mapsto \ D_1 \ \mapsto \ D_2 \ \mapsto \ \cdots \ \mapsto \ D_n =: D'.  \label{eq:seq}
\end{equation}
$D'$ satisfies the MDs as  {\em equality generating dependencies} \cite{Abiteboul}, i.e.
 replacing $\doteq$ by equality.

The semantics specifies the one-step transitions  or updates allowed to go from \bblue{$D_{i-1}$} to \bblue{$D_i$}, i.e. ``$\blue{\mapsto}$" in (\ref{eq:seq}).
Only \bblue{\em modifiable positions} within the instance are allowed to change their values in such a step,
and as forced by the MDs. Actually, the
modifiable positions syntactically depend on a whole set $M$ of MDs and instance at hand; and can be recursively defined (see \cite{front12,datalog12}
for the details). Intuitively, a position $\blue{(t,A)}$ is modifiable iff: (a)
There is a $\blue{t'}$ such that $\blue{t}$ and $\blue{t'}$ satisfy the
similarity condition of an MD with $A$ on the RHS; or (b) $\blue{t[A]}$ has not already been resolved (it is
different from one of its other duplicates).

\begin{example} \  For the schema $R(A,B)$, consider the MD \ \bblue{$R[A] = R[A] \ra R[B]\doteq R[B]$}, and the instance $R(D)$ below.
The positions of the underlined values in $\blue{D}$ are modifiable, because
their values are unresolved (wrt the MD and instance $R(D)$).

\vspace{-2mm}
\begin{center}
\begin{tabular}{l|c|c|}\hline
$R(D)$ & $A$ & $B$ \\ \hline
$t_1$ & $a$ & \underline{$b$} \\
$t_2$ & $a$ & \underline{$c$}\\ \hhline{~--}
\end{tabular}\vspace{1mm}
$~~~\mapsto~~~$
\hspace*{4mm}\begin{tabular}{l|c|c|}\hline
$R(D')$ & $A$ & $B$ \\ \hline
$t_1$ & $a$ & $\blue{d}$ \\
$t_2$ & $a$ & $\blue{d}$\\ \hhline{~--}
\end{tabular}
\end{center}

\vspace{-2mm}
\noindent $\blue{D'}$ is a \bblue{resolved instance} since it satisfies the
MD interpreted as the FD $R: A \rightarrow B$. Here, the update value \bblue{$d$} is arbitrary.

$\blue{D'}$ has no modifiable positions with unresolved values:
the values for $B$ are already the
same, so there is no reason to change them. \boxtheorem
\end{example}

More formally, the {\em single step semantics} ($\mapsto$ in (\ref{eq:seq})) is as follows.
Each pair $\blue{D_i, D_{i+1}}$ in an update
sequence (\ref{eq:seq}), i.e. a chase step, must {\em satisfy} the set $M$ of MDs {\em modulo unmodifiability},
denoted \ $(D_i,D_{i+1}) \models_{\it um} M$, which   holds iff: \ (a)
 For every MD in $M$, say \ $R[\bar A]\approx S[\bar B]\ra$ $R[\bar C]\doteq S[\bar D]$
\ and pair of tuples $\blue{t_R}$ and $\blue{t_S}$,
if $\blue{t_R[\bar A]\approx t_S[\bar B]}$ in $\blue{D_i}$, then $\blue{t_R[\bar C] = t_S[\bar D]}$
in $\blue{D_{i+1}}$; and (b)
 The value of a position can only differ between
$\blue{D_i}$ and $\blue{D_{i+1}}$ if it is modifiable wrt $\blue{D_i}$. Accordingly, in (\ref{eq:seq}) we also require that
$(D_i,D_i) \not \models_{um} M$, for $i < n$, and  $(D_n,D_n) \models_{um} M$ (the {\em stability} condition).\footnote{The case $D' = D_0$ occurs only when
$D$  is already resolved.}

This semantics stays as close as possible to the spirit of
the MDs as originally introduced \cite{Fan09}, and also {\em uncommitted} in the sense that the MDs do not specify
how the matchings have to be realized (c.f. Section \ref{sec:disc} for a discussion).

\begin{example} \label{ex:two} \  Consider the following instance and set of MDs. Here, attribute $\blue{R(C)}$ is changeable. Position \blue{$(t_2,C)$ is not modifiable} wrt. $\blue{M}$
and $\blue{D}$: There is no justification to change its value {\em in one step} on the basis

\vspace{-1mm}
\begin{multicols}{2}
\hspace*{-0.5cm}
\begin{tabular}{l|c|c|c|} \hline
$R(D)$ & $A$ & $B$ & $C$ \\ \hline
$t_1$ &$a$ & $b$ & $d$ \\
$t_2$ &$a$ & $c$ & $\underline{e}$ \\
$t_3$ &$a$ & $b$ & $e$\\ \hhline{~---}
\end{tabular}

\phantom{mmmmmmmm}

\vspace*{-12mm}
\begin{eqnarray}
\hspace*{-10mm}R[A] = R[A] &\ra& R[B]\doteq R[B]\nn\\
\hspace*{-10mm}R[B] = R[B] &\ra& R[C]\doteq R[C]. \nn
\end{eqnarray}
\noindent
\end{multicols}

\vspace{-2mm}
\noindent
  of an MD and $\blue{D}$. However,
position \blue{$(t_1,C)$ is modifiable}. $D$ has two resolved instances,   $\blue{D_1}$ and $\blue{D_2}$.

\vspace{-2mm}
\begin{multicols}{2}
{\small
\begin{center}
\begin{tabular}{l|c|c|c|}\hline
$R(D_1)$ & $A$ & $B$ & $C$ \\ \hline
$t_1$ &$a$ & $b$ & $d$ \\
$t_2$ &$a$ & $b$ & $d$ \\
$t_3$ &$a$ & $b$ & $d$\\ \hhline{~---}
\end{tabular}\vspace{2mm}
~~~~~~\begin{tabular}{l|c|c|c|} \hline
$R(D_2)$ & $A$ & $B$ & $C$\\ \hline
$t_1$ & $a$ & $b$ & $e$\\
$t_2$ & $a$ & $b$ & $e$\\
$t_3$ & $a$ & $b$ & $e$\\ \hhline{~---}
\end{tabular}
\end{center} }

\noindent $\blue{D_1}$ cannot be obtained in a single (one step) update since
the underlined value is for a non-modifiable position. However,  $\blue{D_2}$ can. \boxtheorem
\end{multicols}
\end{example}
For arbitrary sets of MDs, some (admissible) chase sequences may not terminate. However, it can be proved that there are always
terminating chase sequences. As a consequence, for some sets of MDs, there are both terminating and non-terminating chase sequences. In any case, the class of resolved instances is always well-defined.

\begin{example}\label{ex:newex}
Consider relation $R[A,B]$, equality as the similarity relation, and the MDs and instance below:
\vspace{-5mm}
\begin{multicols}{2}
\begin{eqnarray*}
m_1:&&\hspace{-6mm}R[A] = R[A]\ra R[B]\doteq R[B]\\
m_2:&&\hspace{-6mm}R[B] = R[B]\ra R[A]\doteq R[A]
\end{eqnarray*}

\begin{center}
\begin{tabular}{c|c|c|} \hline
$R(D)$ & $A$ & $B$ \\ \hline
$t_1$ & $a$ & $c$ \\
$t_2$ & $b$ & $c$ \\
$t_3$ & $b$ & $d$ \\
$t_4$ & $a$ & $d$ \\ \cline{2-3}
\end{tabular}
\end{center}
\end{multicols}
The chase may not terminate,
which happens when the values oscillate, as in the following update sequence:
\begin{center}
\begin{tabular}{c|c|c|} \hline
$R(D)$ & $A$ & $B$ \\ \hline
$t_1$ & $a$ & $c$ \\
$t_2$ & $b$ & $c$ \\
$t_3$ & $b$ & $d$ \\
$t_4$ & $a$ & $d$ \\ \cline{2-3}
\end{tabular}
~~$\mapsto$~~
\begin{tabular}{c|c|c|} \hline
$R(D)$ & $A$ & $B$ \\ \hline
$t_1$ & $a$ & $c$ \\
$t_2$ & $a$ & $d$ \\
$t_3$ & $b$ & $d$ \\
$t_4$ & $b$ & $c$ \\ \cline{2-3}
\end{tabular}
~~$\mapsto$~~
\begin{tabular}{c|c|c|} \hline
$R(D)$ & $A$ & $B$ \\ \hline
$t_1$ & $a$ & $c$ \\
$t_2$ & $b$ & $c$ \\
$t_3$ & $b$ & $d$ \\
$t_4$ & $a$ & $d$ \\ \cline{2-3}
\end{tabular}
$~~\mapsto \cdots$
\end{center}
However, there are non-trivial  terminating chase sequences:
\begin{center}
\begin{tabular}{c|c|c|} \hline
$R(D)$ & $A$ & $B$ \\ \hline
$t_1$ & $a$ & $c$ \\
$t_2$ & $b$ & $c$ \\
$t_3$ & $b$ & $d$ \\
$t_4$ & $a$ & $d$ \\ \cline{2-3}
\end{tabular}
~~$\mapsto$~~
\begin{tabular}{c|c|c|} \hline
$R(D)$ & $A$ & $B$ \\ \hline
$t_1$ & $f$ & $h$ \\
$t_2$ & $f$ & $g$ \\
$t_3$ & $e$ & $g$ \\
$t_4$ & $e$ & $h$ \\ \cline{2-3}
\end{tabular}~~,~~
\end{center}
with $e$, $f$, $g$, and $h$ arbitrary. After this, a stable instance can be obtained by updating all values
in the $A$ and $B$  to the same value.
\boxtheorem
\end{example}

We prefer {\em resolved instances} that are  the closest to the original instance.
A {\em minimally resolved instance} (MRI) of $\blue{D}$ is a resolved
instance $\blue{D'}$ whose {\em the number of changes of attribute values} wrt. $\blue{D}$  is
a minimum. In Example \ref{ex:two}, instance $\blue{D_2}$ is an MRI, but not $\blue{D_1}$ \ (2 vs. 3 changes).
We denote with $\nit{Res}(D,M)$ and $\nit{MinRes}(D,M)$ the classes of resolved, resp. minimally resolved, instances of $D$
wrt $M$.

Infinite chase sequences may occur when the
MDs cyclically depend on each other, in which case
updated instances in a such a sequence may alternate between two or more states \cite[Example 6]{sum12} (see also Example \ref{ex:newex}). However, for the chase sequences that do terminate
in a minimally resolved instance,
the chase imposes a relatively easily characterizable structure \cite{sum12,datalog12}, allowing us to obtain a query rewriting methodology. So, cycles help us achieve tractability
for some classes of queries \cite{datalog12} (cf. Section \ref{sec:queries}).

On the other side, it has been shown that if a set of MDs satisfies a
certain acyclicity property, then all chase sequences terminate after
a number of iterations that depends only on the set of MDs and not
on the instance \cite[Lemma 1]{front12} (cf. Theorem \ref{thm:boundedchase} below). But the number of resolved instances may still be ``very large". Sets of MDs considered in this work are acyclic.

\subsection{Resolved query answers}\label{sec:queries}

Given a conjunctive query \bblue{$\mc{Q}$}, a set of MDs \bblue{$M$}, and an
instance \bblue{$D$}, the {\em resolved answers} to $\mc{Q}$ from $D$ are invariant under the
entity resolution process, i.e.
they are answers to \bblue{$\mc{Q}$} that are true in all
MRIs of $\blue{D}$:
\begin{eqnarray}
\nit{ResAns}_M(\mc{Q},D)&:=&\{\ \bar{a} \ \ | \ D' \models \mc{Q}[\bar{a}], \mbox{ for every } D' \in \nonumber \\ &&~~~~~~~~~~~~\nit{MinRes}(D,M)\}. \label{eq:answers}
\end{eqnarray}
The corresponding decision problem is \
$\nit{RA}(\mathcal{Q},M) := $ \linebreak $\{(D,\bar{a})~|~ \bar{a} \in \nit{ResAns}_M(\mc{Q},D)\}$.

In \cite{datalog12,sum12}, a query rewriting methodology for {\em resolved query answering} (RQA) under MDs (i.e. computing resolved answers to
queries) was presented. In this case,
the rewritten queries turn out to be Datalog
queries with counting, and can be obtained for two main kinds of sets of MDs: (a)   MDs do not depend on each other, i.e. {\em non-interacting} sets of MDs \cite{front12};
(b) MDs that depend cyclically on each other, e.g. as in the set containing
$R[A]\approx R[A]\ra$ $R[B]\doteq R[B]$ and
$\blue{R[B]\approx R[B]\ra R[A]\doteq R[A]}$ (or relationships like this by transitivity).

For these sets of MDs just mentioned, a conjunctive query can be rewritten
to retrieve, in polynomial time in data, the resolved answers, provided
the queries have no joins on existentially quantified
variables corresponding to changeable attributes.  The latter form the class of {\em unchangeable
attribute join conjunctive}  (\bblue{UJCQ}) queries \cite{sum12}.

For example, for the MD $R[A] = R[A]\ra R[B,C]\doteq R[B,C]$ on schema $R[A,B,C]$,
$\mathcal{Q}\!:
\ex x \ex y \ex z(R(x,y,c)\w R(z,y,d))$ is {\em not} UJCQ; whereas
$\mathcal{Q}'\!:
\ex x \ex z(R(x,y,z)\w R(x,y',z')$ is UJCQ. For queries outside UJCQ, the resolved answer problem can be intractable even for
one MD \cite{sum12}.

The case of a set of MDs consisting of both
\begin{eqnarray}
R[A]\approx R[A]&\ra& R[B]\doteq R[B],\label{eq:lid} \\R[B]\approx R[B]&\ra& R[C]\doteq R[C],\nonumber
\end{eqnarray}
which is neither non-interacting nor cyclic, is
not covered by the positive cases for Datalog rewriting above. Actually, for this set RQA becomes intractable for very
simple queries, like $\mathcal{Q}(x, z)\!: \exists y R(x, y, z)$, that is UJCQ \cite{front12}. Sets of MDs like (\ref{eq:lid})
are the main focus of this work.

\section{Intractability of RQA}\label{sec:new}

In the previous section we briefly described classes of queries and MDs for which RQA can be done in
polynomial time in data (via the Datalog rewriting). We also showed that there are intractable cases, by pointing to a specific query and
set of MDs. Natural questions that we start to address in this section are the following: \ (a) What happens outside the Datalog rewritable cases in terms of complexity of
RQA? \
 (b) Do the exhibited query and MDs correspond to a more general pattern for which intractability holds?

For all sets $M$ of MDs we consider below, {\em we assume that at most two relational predicates, say $R, S$, appear in $M$}, e.g. as in $M = \{
R[A] \approx S[B] \ra R[C]\doteq S[E]\}$. In same cases we assume that there are exactly two predicates. The purpose of
this restriction is to simplify the presentation. All results can be generalized to sets of MDs with more than
two predicates. To do this, definitions and conditions concerning the
two relations in the MDs can be extended to cover the additional relations as well.

At the other extreme, when a single predicate occurs in $M$, say $R$, as in Example \ref{ex:two}, the results for at most two predicates can be  reformulated and applied
by replacing $S$ with $R'$. Although $R$ and $R'$ are the
same relation in this case, the prime is used to distinguish between the two tuples
to which the MD refers.

All the sets of MDs considered below are both interacting (non-interaction does not bring complications) and acyclic. Both notions and others can be captured in terms of the MD {\em graph},
 $\nit{MDG}(M)$, of $M$. It is  a directed graph, such that, for $m_1, m_2 \in M$, there is an edge from $m_1$ to $m_2$ if there is
an overlap between $\nit{RHS}(m_1)$ and $\nit{LHS}(m_2)$ (the right- and left-hand sides of the arrows as sets of attributes) \cite{front12}.
 Accordingly, $M$ {\em is acyclic when $\nit{MDG}(M)$ is acyclic}. In fact, the sets of MDs in this work  satisfy a stronger property, defined below,
 which we call {\em strong acyclicity}.

 \begin{definition}\label{def:atclosure} \cite{front12}
1. \ Let $M$ be a set of MDs on schema $\mc{S}$. \
(a) The symmetric binary relation $\doteq_r$ relates
attributes $R[A]$, $S[B]$ of $\mc{S}$ whenever
there is $m \in M$ in which $R[A]\doteq S[B]$ occurs. \
(b) The {\em attribute closure} of
$M$ is the reflexive and transitive
closure of $\doteq_r$. \
(c)  $E_{R[A]}$  denotes the equivalence class of attribute $R[A]$ in the
attribute closure
of $M$.

\noindent 2. \ The {\em augmented MD-graph} of
$M$, denoted $\nit{A\!MDG}(M)$, is a directed graph with a vertex labeled
with
$m$ for each $m\in M$, and with an edge from
$m$ to $m'$ iff there is an attribute, say $R[A]$, with $R[A]\in \nit{RHS}(m)$
and $E_{R[A]}\cap \nit{LHS}(m')\neq \emptyset$.

\noindent 3. \ $M$ is {\em strongly acyclic} if $\nit{AMDG}(M)$ has no
cycles. \boxtheorem
\end{definition}

\ignore{\begin{definition}\label{def:atclosure} \cite{front12}
Let $M$ be a set of MDs on schema $\mc{S}$.\linebreak
(a) The symmetric binary relation $\doteq_r$ relates
attributes $R[A]$, $S[B]$ of $\mc{S}$ whenever
there is $m \in M$ in which $R[A]\doteq S[B]$ occurs.
\newline
(b) The {\em attribute closure} of
$M$ is the reflexive and transitive
closure of $\doteq_r$. \newline
(c)  $E_{R[A]}$  denotes the equivalence class of attribute $R[A]$ in the
attribute closure
of $M$.\boxtheorem
\end{definition}

\vspace{-5mm}
\begin{definition}\label{def:augmented} \cite{front12}
Let $M$ be a set of MDs. The {\em augmented MD-graph} of
$M$, denoted $\nit{A\!MDG}(M)$, is a directed graph with a vertex labeled
with
$m$ for each $m\in M$, and with an edge from
$m$ to $m'$ iff there is an attribute, say $R[A]$, with $R[A]\in \nit{RHS}(m)$
and $E_{R[A]}\cap \nit{LHS}(m')\neq \emptyset$. \boxtheorem
\end{definition}

\vspace{-5mm}
\begin{definition}\label{def:strongly}
A set of MDs $M$ is {\em strongly acyclic} if there are no
cycles in $\nit{AMDG}(M)$. \boxtheorem
\end{definition}}
Because $R[A]\in E_{R[A]}$, for any set $M$ of MDs, all edges in
$\nit{MDG}(M)$ are also edges in $\nit{A\!MDG}(M)$. Therefore, strong acyclicity
implies acyclicity. However, the converse is not true, as shown in the next example.

\begin{example}\label{ex:stronglyacyclic}
The set $M$ of MDs \vspace{-2mm}
\bea
m_1:~R[F]\approx S[G]\ra R[A]\doteq S[H],\nn\\
m_2:~R[A]\approx S[B]\ra R[C]\doteq S[E],\nn\\
m_3:~R[C]\approx S[E]\ra R[I]\doteq S[H],\vspace{-3mm}\nn
\eea
is acyclic but not strongly acyclic. $\nit{MDG}(M)$ has
three vertices,
$m_1, m_2, m_3$, and edges $(m_1,m_2)$
and $(m_2,m_3)$. $\nit{AMDG}(M)$ has the additional edge
$(m_3,m_2)$, because \linebreak $E_{R[I]} = \{R[I],$
$S[H],R[A]\}\cap \nit{LHS}(m_2) = \{R[A]\}$.
\boxtheorem
\end{example}
In this work, we consider strongly acyclic sets of MDs. In particular, two interesting and common kinds that form large classes of sets $M$ of MDs: {\em linear pairs},
which consist of two MDs such that $\nit{MDG}(M)$ contains a
single edge from one to the other (c.f. Definition \ref{def:linearpair});
and acyclic sets that are
{\em pair-preserving} (c.f. Definition \ref{def:pairpreserving}).
From the definitions of these two kinds of sets of MDs it will follow that they are strongly
acyclic.

\begin{theorem}\label{thm:boundedchase} \cite{front12}
Let $M$ be a strongly acyclic set of MDs on schema $\mc{S}$, and $D$ an instance
for $\mc{S}$. Every sequence
of $M$-based updates to $D$ as in (\ref{eq:seq})
terminates with a resolved instance after
at most $d+1$ steps, where $d$ is the maximum
length of a path in $\nit{A\!MDG}(M)$. \boxtheorem
\end{theorem}
As mentioned previously, the chase can be infinite if the set is not
acyclic. Theorem \ref{thm:boundedchase} only tells us about the chase termination and lengths, but it does not involve the data.
So, it does
not guarantee tractability for RQA, leaving room, in principle, for both tractable and intractable cases.
Actually, it can still be the case that there are exponentially many minimally resolved instances.
A reason for this is that the application of an MD to an instance may produce new similarities among the values of
attributes in $\nit{RHS}(m_1)$ that are not strictly required by the chase, but
result from a particular choice of update values. Such ``accidental similarities"
affect subsequent updates, resulting in exponentially many possible update sequences. This is illustrated in the
next example.

\begin{example}\label{ex:accidental}
Consider the strongly acyclic set $M$:
\bea
R[A]\approx R[A] \ra R[B]\doteq R[B],\nn\\
R[B]\approx R[B] \ra R[C]\doteq R[C].\nn
\eea
When the instance
\begin{center}
\begin{tabular}{l|c|c|c|}\hline
$R(D_1)$ & $A$ & $B$ & $C$ \\ \hline
$t_1$ &$a$ & $m$ & $e$ \\
$t_2$ &$a$ & $d$ & $f$ \\
$t_3$ &$b$ & $c$ & $g$\\
$t_4$ &$b$ & $k$ & $h$\\ \hhline{~---}
\end{tabular}
\end{center}
is updated according to $M$, the sets of value positions
$\{t_1[B],$ $t_2[B]\}$ and $\{t_3[B],t_4[B]\}$
must be merged. One possible update is
\begin{center}
\begin{tabular}{l|c|c|c|}\hline
$R(D_1)$ & $A$ & $B$ & $C$ \\ \hline
$t_1$ &$a$ & $m$ & $e$ \\
$t_2$ &$a$ & $d$ & $f$ \\
$t_3$ &$b$ & $c$ & $g$\\
$t_4$ &$b$ & $k$ & $h$\\ \hhline{~---}
\end{tabular}
$~~~\mapsto~~~$
\begin{tabular}{l|c|c|c|}\hline
$R(D_1)$ & $A$ & $B$ & $C$ \\ \hline
$t_1$ &$a$ & $m$ & $e$ \\
$t_2$ &$a$ & $m$ & $f$ \\
$t_3$ &$b$ & $m$ & $g$\\
$t_4$ &$b$ & $m$ & $h$\\ \hhline{~---}
\end{tabular}
\end{center}
The similarities between the  attribute $B$
values of the top and bottom pairs of tuples
are accidental, because they result from the
choice of update values. In the absence of
accidental similarities, there is only one
possible set of sets of values that are merged
in the second update, namely $\{\{t_1[C],t_2[C]\},$
$\{t_3[C],t_4[C]\}\}$.

Accidental similarities
increase the complexity of query answering over the
instance by adding another possible set of sets of
merged values, $\{\{t_1[C],t_2[C],t_3[C],t_4[C]\}\}$.
More generally, for an instance with $n$ sets
of merged value positions in the $B$ column, the
number of possible sets of sets of value positions
in the $C$ column that are merged in the second update
is $\Omega(2^{n^2})$.
\boxtheorem
\end{example}
 We want to investigate the frontier between tractability and intractability. For this reason, we make the assumption that, {\em for each  similarity relation, $\approx$, there
is an infinite set of mutually dissimilar values}. Actually, without this assumption, the
resolved answer problem becomes immediately tractable for certain similarity operators
(e.g. transitive similarity operators). This is because, for these operators,
the whole class of minimal resolved instances of an instance can be computed in polynomial time.

\begin{proposition}\label{prop:poly}
For strongly acyclic sets of MDs, if the similarity predicates are
transitive and there is no infinite set of mutually dissimilar
values, then the set of minimal resolved instances for a given instance $D$ can be computed in
polynomial time in the size of $D$.
\boxtheorem
\end{proposition}

{\it Proof (sketch)}: By Theorem \ref{thm:boundedchase}, the chase terminates after a number of updates that
is constant in the size of the instance. We claim that, at each step
of the chase, the number of updates that could be made that would
lead to a minimal resolved instance is polynomial in the size of the instance. Thus, all minimal
resolved instances can be computed in polynomial time by exhaustively going through all
possible choices of updates at each step.

To prove the claim, we consider, for a given MD $m$, a conjunct $R[A]\approx S[B]$
appearing to the left of the arrow in $m$. Consider a set of tuples in
$R$ and $S$ whose values are merged by the application of $m$. All tuples
in this set must be in the same equivalence class of the transitive closure
of the binary relation expressed by $R[A]\approx S[B]$. By transitivity
of $\approx$, this means that the values that these tuples take on $R[A]$
or $S[B]$ must similar. Thus, there are at most $b$ sets of tuples
whose values are merged, with $b$ the maximum number of mutually distinct
values in a set.

A minimal resolved instance is obtained by choosing,
for each set of values that are merged by application of $m$,
an update value from a set of values consisting of the union of the set of all the values to be
merged and a maximal set of mutually dissimilar
values. If the update value $v$ is not one of the values to be merged,
then the values are updated to $v^*$, which represents
any value from the equivalence class of $\approx$ to which $v$ belongs.
There are a constant number of sets of merged values, and $O(n)$
possible update values for each set, with $n$ the size of the instance.
This proves the claim.
\boxtheorem

Our next results require some terms and notation that we now introduce.

\begin{definition}\label{def:rquery}  
Let $M$ be a set of MDs with predicates $R$ and $S$.
A {\em changeable attribute query} $\mc{Q}$ is a (conjunctive) query  in UJCQ, containing a conjunct
of the form $R(\bar x)$ or $S(\bar y)$ such that all variables
in the conjunct are free and none occur in another conjunct
of the form $R(\bar x)$ or $S(\bar y)$.
Such
a conjunct is called a {\em join-restricted free occurrence} of
the predicate $R$ or $S$.\boxtheorem
\end{definition}
By definition, the class of {\em changeable attribute queries}\linebreak (CHAQ) is a subclass of UJCQ. Both classes depend on
the set of MDs at hand. For example, for the MDs in (\ref{eq:lid}), $\exists y R(x, y, z) \in \mbox{ UJCQ } \! \smallsetminus \! \mbox{ CHAQ}$, but
$\exists w \exists t(R(x, y, z)\wedge$\linebreak $S(x,w, t)) \in \mbox{ CHAQ}$.
We confine our attention to UJCQ and subsets of it, because, as mentioned in the previous
section, intractability limits the applicability of the duplicate resolution method for queries outside UJCQ.

The requirement that the query contains a join-restricted free occurrence
of $R$ or $S$ eliminates from consideration certain queries in UJCQ for which the resolved answer
problem is immediately tractable. For example, for the MDs in (\ref{eq:lid}), the query $\ex y \ex z R(x,y,z)$ is not  CHAQ, and is tractable simply because it does not return the values of a changeable attribute (the resolved answers are the answers in the usual sense). The restriction on joins simplifies the
analysis while still including many useful queries.

In order to eliminate queries like $\ex y \ex z R(x,y,z)$ wrt. $M$ in (\ref{eq:lid}), CHAQ imposes a strong
condition. Actually, the condition can be weakened, requiring to have {\em at least one} of the variables satisfying the condition in the definition for CHAQ. Weakening the condition
makes the presentation much more complex since a finer interaction with the MDs has to be brought into the picture. (We leave this issue for an extended version.)

\begin{definition}\label{def:hard}
A set $M$ of MDs is {\em hard} if for every
CHAQ $\mathcal{Q}$,  $\nit{RA}(\mathcal{Q},M)$
is \nit{NP}-hard. $M$ is {\em easy} if for every CHAQ $\mathcal{Q}$,  $\nit{RA}(\mathcal{Q},M)$ is in \nit{PTIME}.
\boxtheorem
\end{definition}
Of course, a  set of MDs may not be hard or easy. For the
resolved answer problem, membership of
\nit{NP}  is an open problem. However, for
strongly acyclic sets, the bound on the length
of the chase implies an upper bound of $\Pi^P_2$ \cite[Theorem 5]{front12}.

\ignore{\comlb{OLD. Can we say something about how hard they can be, in general? About NP-completeness? Or above? And with the assumptions/restrictions in this paper? Do we have
an example for which RQA is complete for a class higher that NP?}\\
\comj{OLD. I don't have any completeness results. In all cases in the paper, complexity
is bounded above by the second level of the polynomial hierarchy.}\\
\comlb{We should say this (your last comment) explicitly here, citing the result wherever it is (I do not remember if the cases
in \cite{front12} apply here), or formulating it here. Every reviewer would ask how high we can go in the polynomial hierachy, or
even beyond that.}
}

In the following we give some syntactic conditions that guarantee
hardness for classes of MDs. To state them we need to introduce some useful notions first.

\begin{definition} \ 
Let $m$ be an MD. The symmetric binary
relation $\nit{LRel}(m)$ ($\nit{RRel}(m)$) relates each pair of attributes
$R[A]$ and $R[B]$ such that an atom $R[A]\approx S[B]$ (resp. $R[A]\doteq S[B]$)
appears in $m$. An {\em L-component}
({\em R-component}) of $m$ is an equivalence
class of the reflexive and transitive closure ${\it LRel}(m)^{=+}$ (resp. ${\it RRel}(m)^{=+}$) of
$\nit{LRel}(m)$ (resp. $\nit{RRel}(m)$). \boxtheorem
\end{definition}
\begin{example} For $m\!: \
R[A]\approx S[B]\w R[A]\approx S[C] \ \ra \ R[E]\doteq S[F]\w R[G]\doteq S[H]$, there is only one L-component: \
$\{R[A],S[B],S[C]\}$; and two R-components: \  $\{R[E],S[F]\}$
and $\{R[G],S[H]\}$.
\boxtheorem
\end{example}

\subsection{Hardness of linear pairs of MDs}\label{sec:linpairs}

Most of the results that follow already hold for pairs of MDs, we concentrate on this
case first.

\begin{definition}\label{def:linearpair}
 A set  $M = \{m_1,m_2\}$ of MDs is a  {\em linear pair}, denoted by $(m_1,m_2)$, if its graph $\nit{MDG}(M)$
consists of the vertices $m_1$ and $m_2$ with only
an edge from $m_1$ to $m_2$. \boxtheorem
\end{definition}
First, notice that if $(m_1,m_2)$ is a generic linear pair, with
say
\begin{eqnarray}
m_1\!: \ R[\bar A]\approx_1 S[\bar B]&\ra& R[\bar C]\doteq S[\bar E],  \label{eq:mds}\\
m_2\!: \ R[\bar F]\approx_2 S[\bar G]&\ra& R[\bar H]\doteq S[\bar I], \nonumber
\end{eqnarray}
then, from the definition of the MD graph, it follows that $(R[\bar C] \cup S[\bar E]) \cap (R[\bar F] \cup S[\bar G]) \neq \emptyset$, whereas
$(R[\bar H] \cup S[\bar I])  \cap (R[\bar A] \cup S[\bar B]) = \emptyset$. In the following we have to analyze other different forms of (non-)interaction between the attributes in linear pairs.

\begin{definition}\label{def:equivset} \ 
Let $(m_1,m_2)$ be a linear pair as in (\ref{eq:mds}).
\ignore{\bea
m_1\!:~R[\bar A]\approx_1 S[\bar C]\ra R[\bar E]\doteq S[\bar F],\nn\\
m_2\!:~R[\bar G]\approx_2 S[\bar H]\ra R[\bar I]\doteq S[\bar J].  \nn
\eea}
(a)  $B_R$ is a binary (reflexive and symmetric) relation on
attributes of $R$: \ $(R[U_1],R[U_2]) \in
B_R$ iff $R[U_1]$ and $R[U_2]$ are in the same
R-component of $m_1$ or the same L-component of $m_2$.
Similarly for $B_S$.

\noindent (b) An {\em $R$-equivalent set} ($R$-ES) of attributes of $(m_1,m_2)$
is an equivalence class of $\nit{TC}(B_R)$, the transitive closure of $B_R$, with at least one attribute in the equivalence class
belonging to $\nit{LHS}(m_2)$. The definition of an {\em $S$-equivalent set} ($S$-ES) is similar, with $R$ replaced by $S$.

\noindent (c) An ($R$ or $S$)-ES $E$ of $(m_1,m_2)$ is {\em bounded} if
$E\cap  \nit{LHS}(m_1)$ is non-empty. \boxtheorem
\end{definition}

\begin{example} Consider the schema $R[A,C,F,H,I,M]$,\linebreak $S[B,D,E,G,N]$, and the
linear
pair $(m_1,m_2)$ with:

$m_1\!:~R[A]\approx S[B] \ra R[C]\doteq S[D] \ \w
R[C]\doteq S[E] \ \w$\\
\hspace*{3.5cm} $R[F]\doteq S[G]\w R[H]\doteq S[G]$,

$m_2\!:~R[F]\approx S[E]\w R[I]\approx
S[E] \ \w $\\
\hspace*{1.2cm}$R[A]\approx S[E]\w R[F]\approx S[B] \ \  \ra  \ R[M]\doteq S[N]$.
It holds:

\noindent (a) $B_R(R[F],R[H])$ \ due to the occurrence of $R[F]\doteq S[G]$,
$R[H]\doteq S[G]$.

\noindent (b)
$B_R(R[F],R[I])$ \ due to
$R[F]\approx S[E]$, $R[I]\approx S[E]$.

\noindent (c) $B_R(R[I],R[A])$ \
due to $R[I]\approx S[E]$, $R[A]\approx S[E]$.

\noindent (d) $\{R[A],R[F],R[I],R[H]\}$ is an $R$-ES,
and since \linebreak $\{R[A],R[F],R[I],R[H]\}\cap \nit{LHS}(m_1) = \{R[A]\}\neq \emptyset$, it is also bounded.
\boxtheorem
\end{example}
\begin{theorem}\label{thm:combine} \ 
Let \bblue{$(m_1,m_2)$} be a linear pair, with relational
predicates \bblue{$R$} and \bblue{$S$}. Let
\bblue{$E_R$},
\bblue{$E_S$} be
the sets of $R$-ESs and  $S$-ESs, resp. \
The pair \bblue{$(m_1,m_2)$} is \bblue{hard}
if \ \ \bblue{$\nit{RHS}(m_1) \cap \nit{RHS}(m_2) = \emptyset$}, \ and at least one of
(a) and (b) below holds:
\begin{itemize}
\item[(a)] All of the following hold:
\begin{itemize}
\item[(i)]  \  $\nit{Attr}(R) \cap (\nit{RHS}(m_1)\cap\nit{LHS(m_2))} \neq \emptyset$,
 \item[(ii)] \ There are unbounded ESs in \bblue{$E_R$},
 \item[(iii)] \ For some L-component \bblue{$L$} of $m_1$, \newline \hspace*{15mm}$\nit{Attr}(R) \cap (L \cap \nit{LHS}(m_2)) =\emptyset$.
\end{itemize}
\item[(b)] Same as (a), but with $R$ replaced by $S$. \boxtheorem
\end{itemize}
\end{theorem}
Theorem \ref{thm:combine} says that a linear pair of MDs is hard unless
the syntactic form of the MDs is such that
there is a certain association between changeable attributes in $\nit{LHS}(m_2)$ and attributes
in $\nit{LHS}(m_1)$ as specified by conditions (ii) and (iii).

For pairs of
MDs satisfying the negation of (a)(ii) or that of (a)(iii) (or
the negation of (b)(ii) or that of (b)(iii)) in Theorem \ref{thm:combine}, the
similarities resulting from applying $m_2$ are restricted to a subset of
those that are already present among the values of attributes
in $\nit{LHS}(m_1)$, making the problem tractable.
However, when condition
(ii) or (iii) is satisfied, accidental similarities among the values
of attributes in \linebreak $\nit{RHS}(m_1)$ cannot be passed on to values of
attributes in $\nit{RHS}(m_2)$.

\begin{example} The linear pair $(m_1,m_2)$ with
\bblue{\bea
m_1:~R[A]\approx S[B]\ra R[C]\doteq S[D]\nn\\
m_2:~R[C]\approx S[D]\ra R[E]\doteq S[F]\nn
\eea}
 is hard. In fact, first: $\nit{RHS}(m_1) \cap \nit{RHS}(m_2) = \emptyset$.

Now, it satisfies condition (a):
Condition (a)(i) holds, because $R[C] \in \nit{RHS}(m_1)\cap\nit{LHS(m_2)}$.
Conditions (a)(ii) and (a)(iii) are trivially satisfied, because
there are no attributes of $\nit{LHS}(m_1)$ in
$\nit{LHS}(m_2)$. \boxtheorem
\end{example}
As mentioned above, Theorem \ref{thm:combine} generalizes to
the case of more or fewer than two database predicates. It is easy to verify, for the former case, that if
there are more than two predicates in a linear pair, then there must be exactly three
of them, one of which appears in both MDs. In this case,
hardness is implied by condition (a) in Theorem \ref{thm:combine} alone, with $R$ the predicate in common.

\begin{example}\label{ex:generalization1}
The linear pair with three predicates:
\bea
m_1\!:~R[A]\approx S[B]\ra R[C]\doteq S[E],\nn\\
m_2\!:~R[C]\approx P[B]\ra R[F]\doteq P[G].\nn
\eea
is hard if it satisfies condition (a)
in Theorem \ref{thm:combine}. It does satisfy it:
\begin{itemize}
\item[(i)]  \  $\nit{Attr}(R) \cap (\nit{RHS}(m_1)\cap\nit{LHS(m_2))} = \{R[C]\}$.
 \item[(ii)] \ The ES $\{R[C]\}$ is unbound.
 \item[(iii)] \ Part (iii) holds with $L = \{R[A],$ $S[B]\}$.\boxtheorem
\end{itemize}
\end{example}

For the case with only one predicate $R$ in the linear pair, in order to apply Theorem \ref{thm:combine},
we need to derive from it a special result, Corollary \ref{cor:onepredicate} below. It is obtained by first labeling
the different occurrences of the (same) predicate in $M$, and then generating conditions (four of them, analogous to
(a) and (b) in Theorem \ref{thm:combine}) for the
labeled version, $M'$. When $M'$ satisfies those conditions, the original set $M$ is hard. The  algorithm {\em Conditions} in Table \ref{tab:rewrite} does both the
labeling and the condition generation to be checked on $M'$.
 Notice that, after the labeling, there is still only
one predicate in $M'$. The labeling simply provides a convenient
way to refer to different sets of attributes.
Example \ref{ex:generalization2} demonstrates the use of the
algorithm and the application of the corollary.

{\small
\begin{table}[ht]
\centering
\begin{tabular}{|p{8cm}|}\hline
\textbf{Input}: \ A linear pair $M = (m_1,m_2)$ with a single\\\hspace*{1.3cm} predicate $R$\\
\textbf{Output}: \ Two-MD labeled set $M' = \{m_1',m_2'\}$ and \\ \hspace*{1.7cm}four conditions for $M'$\\
{\bf 1.}~\quad Subscript all occurrences of $R$ in $m_1$ ($m_2$) with\\ \hspace*{0.8cm} 1 (2)\\
{\bf 2.}~\quad Superscript all occurrences of $R$ to the left\\ \hspace*{0.8cm}(right) of $\approx$ or $\doteq$
with 1 (2)\\
{\bf 3.}~\quad \textbf{For} each choice of $X,Y\in\{1,2\}$ generate the\\ \hspace*{8mm} following conditions (i),(ii),(iii):\\
\quad (i) \ $\nit{Attr}(R_1^X)\cap \nit{Attr}(R_2^Y) \cap (\nit{RHS}(m_1')\ \cap$\\ \hspace*{10mm}$\nit{LHS(m_2'))} \neq \emptyset$.\\
\quad (ii) \ There are $R^Y$-equivalent sets that do not\\ \hspace*{10mm}contain attributes in
$\nit{Attr}(R_1^X)\cap\nit{LHS}(m_1')$.\\
\quad (iii) \ For some L-component $L$ of $m_1'$,
$\nit{Attr}(R_1^X) \ \cap$\\ \hspace*{10mm} $ \nit{Attr}(R_2^Y) \cap (L \cap \nit{LHS}(m_2')) =\emptyset$.\\\hline
\end{tabular}
\caption{Algorithm {\em Conditions}}\label{tab:rewrite}
\end{table}
}

\begin{corollary}\label{cor:onepredicate}
A linear pair containing one predicate is hard if it satisfies
$\nit{RHS}(m_1) \cap \nit{RHS}(m_2) = \emptyset$ and at least one
of the four sets of three conditions (i)-(iii) generated by Algorithm {\em Conditions}.
\end{corollary}

\begin{example}\label{ex:generalization2}
Consider the linear pair $M$:
\bea
&& m_1\!:~R[A]\approx R[B]\w R[C]\approx R[E]\ra R[F]\doteq R[G]\nn\\
&& \hspace*{5.5cm}\w R[B]\doteq R[G],\nn\\
&& m_2\!:~R[G]\approx R[H]\w R[B]\approx R[I]\w R[L]\approx R[I]\ra\nn\\
&& \hspace*{5.7cm}R[J]\doteq R[K].\nn
\eea
Algorithm {\em Conditions} produces the following labeling:
\bea
&& m_1'\!:~R_1^1[A]\approx R_1^2[B]\w R_1^1[C]\approx R_1^2[E]\ra \nn\\
&& \hspace*{2.6cm}R_1^1[F]\doteq R_1^2[G]\w R_1^1[B]\doteq R_1^2[G],\nn\\
&& m_2'\!:~R_2^1[G]\approx R_2^2[H]\w R_2^1[B]\approx R_2^2[I]\w \nn\\
&& \hspace*{2.6cm}R_2^1[L]\approx R_2^2[I]\ra R_2^1[J]\doteq R_2^2[K].\nn
\eea
With the above labeling, $R^1$
($R^2$)-equivalent sets can be defined analogously to $R$ ($S$)-equivalent sets
in the two relation case, except that they generally include attributes from two ``relations",
$R_1^1$ and $R_2^1$ ($R_1^2$ and $R_2^2$), instead of one. For example,
in $\{m_1',m_2'\}$, one $R^1$-ES is $\{R_1^1[F],R_1^1[B],R_2^1[B],$ $R_2^1[L]\}$.

The conditions output by {\em Conditions} for the combination $X=1$, $Y=2$ is
the following: (i) $\nit{Attr}(R_1^2)\cap \nit{Attr}(R_2^1) \cap (\nit{RHS}(m_1)\cap\nit{LHS(m_2))} \neq \emptyset$,
(ii) There are $R^1$-equivalent sets that do not contain attributes in $\nit{Attr}(R_1^2)\cap\nit{LHS}(m_1)$, and
(iii) For some L-component \bblue{$L$} of $m_1$, $\nit{Attr}(R_1^2)\cap$ \linebreak $\nit{Attr}(R_2^1)$ $ \cap(L \cap \nit{LHS}(m_2)) =\emptyset$.

These conditions are satisfied by $M'$. In fact, for (i) this set is $\{R_1^2[G]\}$; for (ii) the $R^1$-ES $\{R_2^1[G]\}$ satisfies
the condition; and for (iii) $L = \{R_1^1[C],R_1^2[E]\}$ satisfies the condition.
Thus, by Corollary \ref{cor:onepredicate}, $M$ is hard.
\boxtheorem
\end{example}

\vspace{-6mm}
\begin{example}\label{ex:accidental2} (example \ref{ex:accidental} cont.)
This set $M$ is hard by Corollary \ref{cor:onepredicate}. In fact,
Algorithm {\em Conditions} produces the following labeled set $M'$:
\bea
R_1^1[A]\approx R_1^2[A] \ra R_1^1[B]\doteq R_1^2[B],\nn\\
R_2^1[B]\approx R_2^2[B] \ra R_2^1[C]\doteq R_2^2[C];\nn
\eea
which satisfies the conditions (i)-(iii) for the choice $X=1$, $Y=2$: for
(i) this set is $\{R[B]\}$; for (ii) the $R^1$-ES $\{R_1^1[B],R_2^1[B]\}$ satisfies
the property; and for (iii) we use $L = \{R_1^1[A],R_1^2[A]\}$.

As mentioned in Section \ref{sec:queries}, for the given $M$ and the query
$\mathcal{Q}(x, z)\!: \exists y R(x, y, z)$, RQA is intractable \cite{front12}.
This query is in UJCQ $\smallsetminus$ CHAQ. Now, we have just obtained that RQA, for that $M$, is also
intractable for all CHAQ queries. \boxtheorem
\end{example}

\vspace{-2mm}
\begin{example} \label{ex:tractablecase}
Consider $M$ consisting of
\bea
&&m_1:~R[A]\approx R[A]\ra R[B]\doteq R[B],\nn\\
&&m_2:~R[A]\approx R[A]\w R[B]\approx R[B]\ra R[C]\doteq R[C].\nn
\eea
It does not satisfy the conditions of Theorem \ref{thm:combine} (actually, Corollary \ref{cor:onepredicate}).
The sole L-component of $m_1$ is $\{R[A]\}$, and all
attributes of this set occur in $\nit{LHS}(m_2)$. Actually, the
set is easy, because the non-interacting set
\bea
R[A]\approx R[A]\ra R[B]\doteq R[B],\nn\\
R[A]\approx R[A]\ra R[C]\doteq R[C].\nn
\eea
is equivalent to it in the sense that, for any instance,
the MRIs are the same for either set. This is because applying $m_1$ to the tuples
of $R$ and $S$ results in an instance such that
all pairs of tuples satisfying the first conjunct
to the left of the arrow in $m_2$ satisfy the
entire similarity condition.\boxtheorem
\end{example}

Theorem \ref{thm:combine} gives a syntactic condition for hardness.
It is an important result, because it applies to simple sets of
MDs such as that in Example \ref{ex:accidental} that we expect
to be commonly encountered in practice. Moreover, in Section \ref{sec:hardacyclic},
we use Theorem \ref{thm:combine} to show that similar sets
involving more than two MDs are also hard.

The conditions for hardness in Theorem \ref{thm:combine} are not necessary conditions. Actually, the set of MDs in Example \ref{ex:tractabletransitivecase}
below is hard, but does not satisfy the conditions this theorem.

\section{A dichotomy result}\label{sec:dich}

All syntactic conditions/constructs on attributes above, in particular, the transitive closures on attributes, are
``orthogonal" to semantic properties of the similarity relations. When similarity predicates are transitive, every linear pair not
satisfying the hardness criteria of Theorem \ref{thm:combine} is easy.

\begin{theorem}\label{thm:dichotomy}  Let $(m_1, m_2)$ be a linear pair with \linebreak $\nit{RHS}(m_1) \cap \nit{RHS}(m_2) = \emptyset$.  If the
similarity
operators are transitive, then $(m_1, m_2)$ is either
easy or hard. More precisely, if the conditions of Theorem \ref{thm:combine} hold, $M$ is hard. Otherwise, $M$ is easy. \boxtheorem
\end{theorem}

Theorem \ref{thm:dichotomy} does not hold in general when similarity is not transitive (c.f. Proposition \ref{prop:nontransitive} below). The possibilities for accidental
similarities are reduced by disallowing  that two dissimilar values are similar to a same value. Actually, the complexity of the problem is reduced to
the point where the resolved answer problem becomes tractable.

\ignore{
\comlb{You are saying that, as stated above, that there may still be accidental similarities (you say the ``possibilities are reduced"). If this is true, the question is
*by how much* is the reduction. So much that makes an intractable problem tractable?  This is not said. Check, please.}\\
\comj{I changed the sentence in red.}
}

\ignore{\comlb{Give an intuition here about why transitivity helps. \ignore{ (b) Give an example of application, i.e.
a case that does not satisfy the conditions of Theo. 1. (c) Contrast with the negative example to be given for Theo. 1?}}\\
\comj{I added the paragraph above.}
}

\ignore{\comj{OLD. I put the proof and example below.}\\
\comlb{Is this the complete proof or just a sketch? If the latter, we should have a full proof
in the appendix.}\\
\comj{It is a sketch. I included the complete proof in the appendix.} }

\ignore{
\comlb{In the appendix there is a proof of a *Lemma* 2. Is that supposed to be the proof of this? Check, please.}\\
\comj{Yes, I corrected the label in the appendix.}
}

{\it Proof (sketch)}: As discussed in Section \ref{sec:linpairs}, intractability occurs
as a result of the effect of particular choices of update value on subsequent updates.
Obviously, if condition (a)(i) ((b)(i)) of Theorem \ref{thm:combine} does not hold, then changes to values
in $R$ ($S$) in the first update cannot affect subsequent updates. If operators
are transitive and (ii) or (iii) hold, then the effect is sufficiently restricted
that the set of MDs becomes easy.

To illustrate, we will consider updates when condition (a)(iii) does not
hold. Let $m_1$ and $m_2$ be as in Theorem \ref{thm:combine}.
Let $A$ be the set of sets of tuples in $R$ whose values are merged as a
result of applying $m_1$. Let $B$ be the set of sets of tuples in $R$ whose
values are merged as a result of applying $m_2$ in the second update.
We claim that, for any $B_1\in B$, there is at most one $A_1\in A$
such that $A_1\bigcap B_1\neq\emptyset$. This implies that no accidental
similarity between updated values can affect subsequent updates
as in Example \ref{ex:accidental}, from which it follows that the set of MDs is easy.

Let $L$ be an L-component of $m_1$.
To prove the claim, we first prove that,
for any attribute $E\in L$, if a pair of tuples $t_1$ and $t_2$ in $R$
whose values are modified by application of $m_1$
satisfies $t_1[E] \approx t_2[E]$, then $t_1[E] \approx t_2[E]$
for all $E\in L$.
Suppose $t_1[E] \approx t_2[E]$. Since $t_1$ and $t_2$
are modified by $m_1$, there must be tuples $t_3$ and $t_4$ in $S$ such that
the pair $t_1$, $t_4$ and the pair $t_2$, $t_3$ satisfy the similarity condition
of $m_1$. Let $F$ be an attribute of $S$ such that $R[E]\approx S[F]$
is a conjunct of $m_1$. By $t_1[E] \approx t_2[E]$ and transitivity of $\approx$,
$t_3[F] \approx t_4[F]$ holds. More generally, for any pair of attributes
$R[E']$ and $S[F']$ such that $R[E']\approx S[F']$ is a conjunct of $m_1$,
$t_1[E'] \approx t_2[E']$ iff $t_3[F'] \approx t_4[F']$. It then follows
from the definition of L-component that $t_1[E] \approx t_2[E]$ for all
$E\in L$.

Suppose that the values of a pair of tuples $t_1$ and $t_2$ in $R$ are merged
by application of $m_2$ in the second update. By an argument similar to
the preceding, this means that $t_1[A]\approx t_2[A]$ for
any attribute $A$ of $R$ to the left of the arrow in $m_2$.
Since (a)(iii) does not hold, by the result of the preceding paragraph,
$t_1$ and $t_2$ satisfy the similarity condition of $m_1$. This proves the
claim.
\boxtheorem

\begin{example} \label{ex:tractabletransitivecase}
The linear pair $M$ consisting of
\bea
&&m_1:~R[A]\approx S[B]\w R[I]\approx S[J]\ra R[E]\doteq S[F],\nn\\
&&m_2:~R[E]\approx S[F]\w R[A]\approx S[J]\w R[I]\approx S[B]\ra\nn\\
&& \hspace*{5.8cm}R[G]\doteq S[H].\nn
\eea
does not satisfy the conditions of Theorem \ref{thm:combine}, because $m_1$ has two L-components,
$\{R[A],S[B]\}$ and $\{R[I],S[J]\}$. Since $\nit{LHS}(m_2)$
includes one attribute of $R$ and $S$ from each of these L-components,
conditions (a)(iii) and (b)(iii) are not satisfied. Then, by Theorem \ref{thm:dichotomy},  $M$ is easy when $\approx$ is
transitive. \boxtheorem
\end{example}
In Example \ref{ex:tractablecase}, we showed that a pair
of MDs is easy for arbitrary $\approx$ by exhibiting an equivalent non-interacting
set. This method cannot be applied in Example \ref{ex:tractabletransitivecase}, because
the similarity condition of $m_1$ is not included in that
of $m_2$. Actually, the set of MDs
in Example \ref{ex:tractabletransitivecase} can be hard for non-transitive similarity relations, as the following
proposition shows.

\begin{proposition}\label{prop:nontransitive}
There exist (non-transitive) similarity operators $\approx$ for which
the set of MDs in Example \ref{ex:tractabletransitivecase}
is hard.
\boxtheorem
\end{proposition}

\section{Hardness of acyclic sets of MDs}\label{sec:hardacyclic}

We consider now acyclic sets of MDs of arbitrary finite size, concentrating on a class of them that
is common in practice.

\begin{definition}\label{def:pairpreserving}
A set $\blue{M}$  of MDs is {\em pair-preserving} if for every attribute appearing in $M$, say $R[A]$,
there is exactly one attribute appearing in $M$, say $S[B]$, such that $R[A]\approx S[B]$
or $R[A]\doteq S[B]$ (or the other way around) occurs in $M$.
\boxtheorem
\end{definition}
It is easy to verify that pair-preserving, acyclic sets of MDs are strongly acyclic.

\begin{example}\label{ex:pairpreserving}
$M$ in Example \ref{ex:tractablecase}
is pair-preserving. However, the set of MDs in
Example \ref{ex:tractabletransitivecase} is
not pair-preserving, because $S[B]$ is paired
with both $R[A]$ and $R[C]$ in $m_1$. It is also
possible for cyclic sets of MDs to be pair-preserving.
For example, the set
\bea
R[A]\approx R[A]\ra R[B]\doteq R[B],\nn\\
R[B]\approx R[B]\ra R[A]\doteq R[A],\nn
\eea
is pair-preserving.
\boxtheorem
\end{example}
Pair-preservation typically holds in entity resolution, because the values of pairs of attributes are normally
compared only if they hold the same kind of information
(e.g. both addresses or both names).

Now, recall from the previous section that syntactic conditions on linear pairs $(m_1,m_2)$
imply hardness.
One of the requirements is the absence of certain
attributes in $\nit{LHS}(m_1)$ from $\nit{LHS}(m_2)$ (c.f.
conditions (a)(iii) or (b)(iii)).
The condition of {\em
non-inclusiveness} wrt subsets of $M$ is a syntactic condition on acyclic, pair-preserving sets $\blue{M}$ of MDs that generalizes the conditions that ensure hardness
for linear pairs.

\begin{definition}\label{def:inclusion} \ 
Let $M$ be  acyclic and pair-preserving,
$B$ an attribute in $M$, and
$M' \subseteq M$. \  $B$ is {\em non-inclusive}
wrt. $M'$ if, for every $m \in M \! \smallsetminus \! M'$
with $B\in \nit{RHS}(m)$, there is an
attribute $C$ such that: \
(a) $C\in \nit{LHS}(m)$, \
(b) $C\nin \bigcup_{m'\in M'}\nit{LHS}(m')$, and (c) $C$ is {\em non-inclusive}
wrt. $M'$. \  \ignore{An attribute that is not non-inclusive wrt. $M'$ is
{\em inclusive} wrt. $M'$.}
\boxtheorem
\end{definition}
This is a recursive definition of non-inclusiveness. The base case occurs when $C$ is not in $\nit{RHS}(m)$
for any $m$, and so must be inclusive (i.e. not non-inclusive). Because $C\in \nit{LHS}(m)$ in the definition, for any $m_1$ such that $C\in \nit{RHS}(m_1)$, there is an edge from $m_1$ to $m$. Therefore, we are traversing an edge backwards with each recursive step, and the recursion terminates by the acyclicity assumption.

\begin{example}\label{ex:recursivedef}
In the set acyclic and pair-preserving set of MDs containing
\bea
&&m_1:~R[I]\approx S[J]\ra R[A]\doteq S[E],\nn\\
&&m_2:~R[A]\approx S[E]\ra R[C]\doteq S[B],\nn\\
&&m_3:~R[G]\approx S[H]\ra R[I]\doteq S[J],\nn
\eea
$R[A]$ is non-inclusive wrt. $\{m_2\}$ because $R[A]\in \nit{RHS}(m_1)$
and there is an attribute, $R[I]$, in $\nit{LHS}(m_1)$
that satisfies conditions (a), (b), and (c) of Definition
\ref{def:inclusion}. Conditions (a) and (b) are obviously
satisfied. Condition (c) is satisfied, because $R[G]$ is non-inclusive wrt. $\{m_1\}$. This is trivially true,
since $R[G]\nin \nit{RHS}(m_1)\cup \nit{RHS}(m_3)$.
\boxtheorem
\end{example}
Non-inclusiveness is a generalization of conditions (a) (iii) and (b) (iii)
in Theorem \ref{thm:combine} to a set of arbitrarily many MDs. It expresses
a condition of inclusion of attributes in the left-hand side of one MD
in the left-hand side of another. In particular,
suppose $M = (m_1,m_2)$ is a pair-preserving linear pair, and take $M'=\{m_2\}$.
It is easy to verify that the requirement that there is an
attribute in $\nit{RHS}(m_1)$ that is non-inclusive wrt. $M'$ is equivalent to
conditions (a)(iii) and (b)(iii) of Theorem \ref{thm:combine}.

Theorem \ref{thm:main} tells us that
a set of MDs that is non-inclusive in this sense is hard.

\begin{theorem}\label{thm:main} \ 
Let $M$ be acyclic and pair-preserving. Assume
there is $\{m_1,m_2\}$ $\subseteq  M$, and attributes $C\in\nit{RHS}(m_2)$,
$B\in\nit{RHS}(m_1)\bigcap\nit{LHS}(m_2)$ with: \ (a) $C$ is non-inclusive
wrt $\{m_1,m_2\}$, and (b) \ $B$ is non-inclusive wrt $\{m_2\}$.
Then, $M$ is hard. \boxtheorem
\end{theorem}
\begin{example}\label{ex:maintheorem} (example \ref{ex:recursivedef} cont.)
The set of MDs
is hard. This follows from
Theorem \ref{thm:main}, with $m_1, m_2$ in the theorem being the $m_1, m_2$ in the example.
$C, B$ in the theorem are $R[C], R[A]$ in the example, resp. Part
(b) of the theorem was shown in the first part of this example.
Part (a) holds trivially, since $R[C]\nin \nit{RHS}(m_3)$. \boxtheorem
\end{example}

\begin{example}\label{ex:newEx}
Consider $M = \{m_1,m_2,m_3\}$ with
\bea
&&m_1:~R[G]\approx S[H]\ra R[I]\doteq S[J],\nn\\
&&m_2:~R[G]\approx S[H]\w R[I]\approx S[J]\ra R[A]\doteq S[E],\nn\\
&&m_3:~R[G]\approx S[H]\w R[A]\approx S[E]\ra R[C]\doteq S[B].\nn
\eea
It does not satisfy the condition of Theorem \ref{thm:main}.
The only candidates for $m_1$ and $m_2$ in
the theorem are $m_1$ and $m_2$, respectively, and $m_2$ and $m_3$, respectively,
because of the requirement that $\nit{RHS}(m_1)\bigcap\nit{LHS}(m_2)\neq \emptyset$.
In the first case, $B$ in the theorem is $R[I]$ (or $S[J]$), which does not satisfy (b)
because $\nit{LHS}(m_1)\backslash\nit{LHS}(m_2) = \emptyset$.
In the second case, $B$ in the theorem is $R[A]$ (or $S[E]$). Because $R[G]$ and
$S[H]$ are in $\nit{LHS}(m_3)$, $R[A]$ can only satisfy (b) if $R[I]$ does.
$R[I]$ does not satisfy (b), since $\nit{LHS}(m_1)\backslash\nit{LHS}(m_3) = \emptyset$.

Actually, $M$ is easy, because it is equivalent
to the non-interacting set
\bea
&&m_1':~R[G]\approx S[H]\ra R[I]\doteq S[J],\nn\\
&&m_2':~R[G]\approx S[H]\ra R[A]\doteq S[E],\nn\\
&&m_3':~R[G]\approx S[H]\ra R[C]\doteq S[B],\nn
\eea
which can be shown with the same argument as in Example \ref{ex:tractablecase} to $m_1$ and $m_2$, and then to $m_2$ and $m_3$.
\boxtheorem
\end{example}
Our dichotomy result applies to linear pairs (and transitive similarities). However, tractability
can be obtained in some cases of larger sets of MDs for which hardness cannot be obtained via Theorem \ref{thm:main} (because the conditions do not hold).
The following is a general result concerning sets such as
$M$ in Example \ref{ex:newEx}.

\begin{theorem}\label{lem:tractable1}
Let $M$ be an acyclic, pair-preserving set of MDs.
If for all $m\in M$, all changeable attributes $A$ such that $A\in \nit{LHS}(m)$
are inclusive wrt $\{m\}$, then $M$ is easy.
\boxtheorem
\end{theorem}

{\it Proof}: Consider the MD graph $\nit{MDG}(M)$ of $M$.
We transform $M$ to an equivalent set of MDs $M'$ as follows.
For each MD $m$ such that its corresponding vertex $v(m)$ in $\nit{MDG}(M)$ has an incoming edge from a vertex $v'(m')$ that has no incoming edges incident on it, we delete from $\nit{LHS}(m)$ all attributes in $\nit{RHS}(m')$. It is readily verified that the maximum length of a path in $\nit{MDG}(M')$ is one less than the
maximum length of a path in $\nit{MDG}(M)$, and that $M'$ satisfies the conditions of the theorem. Therefore, the transformation can be applied repeatedly until an equivalent
non-interacting set of MDs is obtained.
\boxtheorem

\begin{example}\label{ex:pres2} (example \ref{ex:newEx}
 cont.)
As expected, the set $M$ of MDs $\{m_1,m_2,m_3\}$
 satisfies the requirement of
Theorem \ref{lem:tractable1}.

To show this, the only attributes to be
tested for inclusiveness wrt an MD are $R[A]$ and $R[I]$.
Specifically, it must be determined whether
$R[I]$ is inclusive wrt $\{m_2\}$ and whether $R[A]$ is inclusive
wrt $\{m_3\}$. $R[I]$ is inclusive wrt $\{m_2\}$, because all
attributes in $\nit{LHS}(m_1)$ are in $\nit{LHS}(m_2)$.
$R[A]$ is inclusive wrt $\{m_3\}$, since $R[G]\in \nit{LHS}(m_3)$
and $R[I]$ is inclusive wrt $\{m_3\}$.\boxtheorem
\end{example}

\begin{example} (example \ref{ex:maintheorem} cont.)
The set $\{m_1,m_2,m_3\}$ in Example \ref{ex:recursivedef}
 was shown to
be hard in Example \ref{ex:maintheorem}.

As expected,
it does not satisfy the requirement of Theorem \ref{lem:tractable1}.
This is because $R[A]$ is changeable, $R[A]\in \nit{LHS}(m_2)$, and
$R[B]$ is non-inclusive wrt $\{m_2\}$ since $R[I]\in \nit{LHS}(m_1)$,
$R[I]\nin\nit{LHS}(m_2)$, and $R[I]$ is non-inclusive wrt $\{m_2\}$.
\boxtheorem
\end{example}
As expected, the conditions of Theorems \ref{thm:main} and  \ref{lem:tractable1}
 are mutually exclusive. In fact, $B$ in
Theorem \ref{thm:main} is changeable (since $B\in \nit{RHS}(m_1)$),
$B\in \nit{LHS}(m_2)$, and $B$ is non-inclusive wrt $\{m_2\}$. However, together they do not provide a
dichotomy result, as the following example shows.

\begin{example}\label{ex:another}
The set
\bea
&&m_1:~R[E]\approx R[E]\ra R[B]\doteq R[B]\nn\\
&&m_2:~R[B]\approx R[B]\ra R[C]\doteq R[C]\nn\\
&&m_3:~R[E]\approx R[E]\ra R[C]\doteq R[C]\nn
\eea
does not satisfy the conditions of Theorems \ref{thm:main} or
\ref{lem:tractable1}. It does not satisfy
the condition of Theorem \ref{lem:tractable1} because $R[B]$ is
changeable and non-inclusive wrt $\{m_2\}$. It does not satisfy condition
(a) of Theorem \ref{thm:main}, because $C$ is inclusive wrt $\{m_1,m_2\}$
($R[E]\in \nit{LHS}(m_1)$).

Although tractability of this case cannot be determined through
the theorems above, it can be shown that the set is easy.
The reason is that, for any update sequence
that leads to an MRI, each set of merged duplicates
must be updated to a value in the set (to satisfy
minimality of change). It is easily verified that,
with this restriction, the second update to the
values of $R[C]$ is subsumed by the first, and
therefore this update has no effect on the instance. Thus,
sets of duplicates can be computed in the same way
as with non-interacting sets.\boxtheorem
\end{example}
Notice that
the condition of Theorem \ref{thm:combine} that there exists an ES that is not bounded
does not appear in Theorem \ref{thm:main}. This is because, for pair-preserving, acyclic sets
of MDs, this condition is always
satisfied by any subset of the set that is a linear pair. Indeed,
consider such a subset $(m_1,m_2)$. If all ESs are bounded for this
pair, then by the pair-preserving requirement, $\nit{LHS}(m_2)\subseteq\nit{LHS}(m_1)$.
Since $(m_1,m_2)$ is a linear pair, $\nit{LHS}(m_2)\cap\nit{RHS}(m_1)\neq\emptyset$.
This implies $\nit{LHS}(m_1)\cap$ $\nit{RHS}(m_1)\neq\emptyset$, contradicting
the acyclicity assumption.

For linear pairs, Theorem \ref{thm:main} becomes Theorem \ref{thm:combine}.
For such pairs, condition (a) of Theorem \ref{thm:main} is always satisfied.
If the (acyclic) linear pair is also a pair-preserving, as required by
Theorem \ref{thm:main}, the conditions of Theorem \ref{thm:combine} reduce
to conditions (a)(iii) and (b)(iii), which, as noted previously, are
equivalent to condition (b) of Theorem \ref{thm:main}.

\section{Discussion and Conclusions}\label{sec:disc}

In this paper we have shown that  resolved query
answering is typically intractable when the MDs have a
\bblue{non-cyclic dependence} on each other.

The results in this paper shed additional light on the complexity landscape of resolved query answering
under MDs, complementing previously known results. Actually,
Table 2 summarizes the current state of knowledge of the complexity of the resolved answer problem.

\begin{table*}
\label{tab:first}
\centering
\begin{tabular}{|c|c|c||c|}\hline
\multicolumn{3}{|c||}{\hspace*{-6.5cm}Kind  of MD Set} &
Data Complexity$^\star$\\ \hline\hline
 &
does not satisfy & transitive $\approx$ & easy (Thm. \ref{thm:dichotomy})\\ \cline{3-4}
linear & condition of Thm. \ref{thm:combine} & non-transitive $\approx$
& hard for some $\approx \ ^{\star\star}$    \\
pair & & &  \\ \cline{2-4}
 & satisfies condition & $\nit{RHS}(m_1)\bigcap\nit{RHS}(m_2)= \ems$
& hard (Thm. \ref{thm:combine})\\ \cline{3-4}
 & of Thm. \ref{thm:combine} & $\nit{RHS}(m_1)\bigcap\nit{RHS}(m_2)\neq \ems$
& no general result\\ \hline\hline
 & \multicolumn{2}{|c||}{does not satisfy condition (b) of Thm. \ref{thm:main}} &
easy \\ \cline{2-4}
 &  & satisfies condition  &
hard (Thm. \ref{thm:main})\\
acyclic, pair-& satisfies condition& (a) of Thm. \ref{thm:main} & \\ \cline{3-4}
preserving & (b) of Thm. \ref{thm:main} & does not satisfy  &
can be easy $^{\star\star\star}$ \\
& & condition (a) of Thm. \ref{thm:main} & \\ \hline\hline
\multicolumn{3}{|l||}{\quad HSC (cyclic)}  & easy \cite{sum12}\\ \hline
\end{tabular}

\vspace{1mm}
\hspace*{-4cm}$\star$: \ for all the sets of MDs below, $\Pi^P_2$ is an upper bound

\hspace*{2mm}$\star\star$: \ there are non-transitive similarities for which the set is hard (no known easy set)

$\star\star\star$: \ tractable for some queries (may be intractable for others, but no example known)

\vspace{-3mm}
\caption{Complexity of RQA for Sets of MDs (CHAQ queries)}
\end{table*}

The definition of resolved answer is reminiscent of that of consistent query answer (CQA) in databases that may not
satisfy given integrity constraints (ICs) \cite{Arenas99,B2006}. Much research in CQA has been about developing (polynomial-time) query rewriting methodologies.
The idea is to rewrite a query, say conjunctive, into a new query such that the new query on the inconsistent database returns as usual answers the consistent
answers to the original query.

In all the cases identified in the literature on CQA (see \cite{bertossi11,foiks14} for recent surveys) depending on the class of conjunctive
query and ICs involved, the rewritings that produce polynomial time CQA have been first-order. For MDs, the exhibited rewritings that can be evaluated in polynomial time are in Datalog \cite{datalog12}.

Resolved query answering under MDs brings many new challenges in comparison to CQA, and results for the latter cannot be applied (at least not in an obvious manner): (a)  MDs contain the usually non-transitive
similarity relations. (b) Enforcing consistency of updates requires computing the transitive
closure of such relations. (c) The minimality of {\em value changes} that is not always used in CQA or considered for consistent rewritings. Actually, {\em tuple-based} repairs are usually considered
in CQA \cite{bertossi11}. (d) The semantics of resolved query answering for MD-based
entity resolution
is given, in the end, in terms of a chase procedure.\footnote{For some implicit connections between repairs and chase procedures, e.g. as used in data exchange see
\cite{kolaitisICDT12}, and as used under database completion with ICs see \cite{cali03}.} However, the semantics of CQA is model-theoretic, given in terms repairs that are not operationally defined, but arise
from set-theoretic conditions.\footnote{For additional discussions of differences and connections between CQA and resolved query answering see  \cite{front12,sum12}.}

In this paper we have presented the first dichotomy result for the complexity of
resolved query answering. The cases for this dichotomy depend on the set of MDs, for
a fixed class of queries. In CQA with functional dependencies, dichotomy results have been obtained for limited
classes of conjunctive queries \cite{pema,maslowski,wijsenPods13,koutris}. However, in CQA the cases depend mainly on the queries, as opposed to the FDs.

Some open problems that are subject to ongoing research are about: (a) Obtaining tighter upper-bounds on the complexity of resolved query answering. \ (b) Extending the class of CHAQ
queries, considering additional projections, and also boolean queries. \ (c) Since a condition for easiness was presented
for linear pairs with transitive similarity, deriving
similar results for other commonly used similarity relations, e.g. edit distance. \ (d) Deriving a dichotomy result for acyclic, pair-preserving sets analogous
to the one for linear pairs. \ (e) Since, functional dependencies (and other equality generating dependencies) can be expressed as MDs, with equality
as a transitive symmetry relation, applying the dichotomy result in Theorem \ref{thm:dichotomy} to CQA under FDs (EGDs) under a  {\em value-based} repair
semantics \cite{bertossi11}.

The results
in this paper depend on the chase-based semantics for clean instances that was introduced in Section \ref{sec:sem}. Alternative semantics for clean instances in relation to the
chase sequence in (\ref{eq:seq}) can be investigated \cite{semantics}.\footnote{In \cite{icdt11,tocs,kr12} a chase-based semantics that applies one MD at a
time and uses {\em matching functions} to choose a value for a match has been developed. The introduction of matching functions changes basically the whole picture.}
A couple of them are essentially as follows:\footnote{For more details, see \cite{draft}.}
\begin{itemize}
\item [(a)] Apply a chase that, instead of applying all the MDs, applies only one MD at a time.
\item [(b)] Apply a chase (as in Section \ref{sec:sem}), but making sure  that previous resolutions are never  unresolved later in the process.
\end{itemize}
In  case (b) above, the
same rewriting techniques of \cite{datalog12} apply, but now
also to some sets of MDs with non-cyclic dependencies.

Still in case (b) and acyclic pairs of MDs, we may obtain a different behavior wrt the semantics used in this work.
For example, the resolved query answer problem for $M$ consisting of
\begin{eqnarray}
m_1:~R[A]\approx R[A]\ra R[B]\doteq R[B],\label{MD1}\\
m_2:~R[B]\approx R[B]\ra R[C]\doteq R[C],\nonumber \label{MD2}
\end{eqnarray}
was established as hard in Example \ref{ex:accidental2}. However, under the semantics in (b), it becomes
tractable for every UJCQ \cite{semantics}. The reason is that,
while accidental similarities can arise among values of $R[B]$
in the update process, these similarities cannot affect subsequent
updates to values of $R[C]$ (c.f. Example \ref{ex:accidental}).
If a pair of tuples must have their $R[C]$ attribute values merged
in the second update as a result of an accidental similarity between their $R[B]$
values, these values would have to be merged anyway, to preserve
similarities generated in the $R[C]$ column by the first update.

In a different direction, even with the semantics used in this work (as in Section \ref{sec:sem}), we could consider an alternative definition of resolved answer to the
one given in (\ref{eq:answers}), namely
those that are true in {\em all}, not necessarily minimal, resolved instances, i.e. in the instances in $\nit{Res}(D,M)$ (as opposed
to $\nit{MinRes}(D,M)$),
obtaining a subset of the original resolved answers. For some sets of MDs, like the one in
(\ref{MD1}) above, the different possible sets
of merged positions in resolved instances (not directly the resolved instances though) can be specified in
(extensions of) Datalog.\footnote{This does not extend to minimally resolved instances since the sets of merged positions may not coincide
with those for general resolved instances.}
These rules can be combined with a query to produce a new query that
retrieves the resolved answers under this alternative query answer semantics \cite{semantics}.

\vspace{2mm}
\noindent {\bf Acknowledgments:} \ Research supported by the NSERC Strategic
Network on Business Intelligence (BIN ADC05), NSERC/IBM CRDPJ/371084-2008, and NSERC Discovery.

\newpage
\appendix

\section{Proofs of Results} \label{sec:proofs}

For the proofs below, we need some auxiliary definitions and results.

\begin{definition}\label{def:closure}
Let $m$ be an MD. Consider the binary relation
that relates pairs of tuples that satisfy the
similarity condition of $m$. We denote the
transitive closure of this relation by $T_m$.
\boxtheorem
\end{definition}

The relation $T_m$ is an equivalence relation, since
reflexivity and symmetry are satisfied by the relation
of which it is the transitive closure.

\begin{lemma}\label{lem:tc}
Let $D$ be an instance and let
$m$ be an MD.
 An instance $D'$
satisfies $(D,D')\models_{\it um} m$ iff
for each equivalence
class of $T_m$, for tuples $t_1$ and $t_2$
in the equivalence class, and for attributes
$A$ and $B$ in the same R-component of $m$,
it holds that $t_1'[A] = t_2'[B]$, where
$t_1'$ ($t_2'$) is the tuple in $D'$ with the same
identifier as $t_1$ ($t_2$).
\end{lemma}
{\em Proof}: Suppose $(D,D')\models_{\it um} m$, and let
$t_1$, $t_2$, A, and B be as in the statement of the
theorem. The tuples
$t_1'[A]$ and $t_2'[B]$ are equivalent under the
equivalence relation obtained by taking the transitive closure of
the equality relation. But the equality relation is its own
transitive closure. Therefore, $t_1'[A] = t_2'[B]$.
The converse is trivial.
\boxtheorem

\begin{definition}\label{def:coverset}
Let $S$ be a set and let $S_1$, $S_2$,...$S_n$ be
subsets of $S$ whose union is $S$. A {\em cover subset} is a subset
$S_i$, $1\leq i\leq n$, that is in a smallest
subset of $\{S_1,S_2,...S_n\}$ whose union is $S$.
The problem {\em Cover Subset (CS)} is the problem
of deciding, given a set $S$, a set of subsets
$\{S_1,S_2,...S_n\}$ of $S$, and an subset
$S_i$, $1\leq i\leq n$, whether or not $S_i$
is a cover subset.\boxtheorem
\end{definition}

\begin{lemma}\label{lem:minset}
CS and its complement are $\nit{NP}$-hard.
\end{lemma}
{\em Proof}: The proof is by Turing reduction from
the minimum set cover problem, which is $\nit{NP}$-complete.
Let $O$ be an oracle for CS.
Given an instance of minimum set cover consisting
of set $S$, subsets $S_1$, $S_2$,...$S_n$ of $S$,
and integer $k$, the following algorithm determines
whether or not there exists a cover of $S$ of size
$k$ or less. The algorithm queries $O$ on $(S,\{S_1,...S_n\},S_i)$
until a subset $S_i$ is found for which $O$ answers
yes. The
algorithm then invokes itself recursively on the
instance consisting of set $S\backslash S_i$,
subsets \\$\{S_1,...S_{i-1},S_{i+1},...S_n\}$, and
integer $k-1$. If the input set in a recursive call
is empty, the algorithm halts and returns yes, and
if the input integer is zero but the set is nonempty,
the algorithm halts and returns no. It can be shown
using induction on $k$ that this algorithm returns the
correct answer. This shows that CS is $\nit{NP}$-hard.
The complement of CS is hard by a similar proof, with the oracle for
CS replaced by  an oracle for the complement of CS.
\boxtheorem

\vspace{2mm}
\defproof{Theorem \ref{thm:combine}}{
For simplicity of the presentation,
we make the assumption that, for relations
$R$ and $S$, the domain of all
attributes that occur in $m_1$ and $m_2$ is the same. If this assumption
does not hold, the general form of the instance
produced by the reduction would be the same, but
it would have different sets of values for attributes with different
domains. All pairs of
distinct values in an instance are dissimilar. Unless otherwise
noted, when we refer to the equivalence classes of
$T_{m_1}$ or $T_{m_2}$, we mean the non-singleton equivalence
classes of these relations.

Wlog, we will assume that part (a) of
Theorem \ref{thm:combine} does not hold.
A symmetric argument proves the theorem for the
case in which (b) does not hold.
Let $E$ and $L$ denote an ES and
an L-component that violate
part (a) of Theorem \ref{thm:combine}.
 We prove the theorem separately for the following
three cases: (1) There exists such an $E$ that contains only attributes of
$m_1$, (2) there exists such an
$E$ that contains both attributes not
in $m_1$ and attributes in $m_1$, and (3) (1) and (2) don't hold
(so there exists such an $E$ that contains only attributes
not in $m_1$). Case (1) is divided
into two subcases: (1)(a) Only one
R-component of $m_1$ contains attributes of
$E$ and (1)(b) more than one R-component
contains attributes of $E$.

In addition to the constants that are introduced
for each case, we introduce a constant
$c_d$ from each attribute domain $d$.
$R$ ($S$) contains a tuple
that takes the value $c_d$ on each attribute
of $R$ ($S$), where $d$ is the domain of the
attribute. For all other tuples besides this one,
the values of attributes not in $m_1$ or $m_2$
are arbitrary for the instance produced by the
reduction.

 For any relation $W$
other than $R$ and $S$, the tuples that are contained
in $W$ are specified in terms of those contained
in $R$ and $S$ as follows. Let $X$ be the set of
attributes of $W$ whose domain is the same as that
of the attributes in $m_1$ and $m_2$, and let $Y$ be the
set of all other attributes of $W$. $W$ contains the
set of all tuples such that, for each attribute
in $Y$, the attribute takes the value $c_d$, where
$d$ is the domain of the attribute, and for each
attribute in $X$, the attribute takes a value
of an attribute in $m_1$ or $m_2$.

Case (1)(a):  We reduce an instance of the compliment of CS
(c.f. Definition \ref{def:coverset})
to this case, which is $\nit{NP}$-hard by lemma \ref{lem:minset}.
Let $F$ be an instance of CS with
set of elements $U = \{e_1,e_2,...e_n\}$ and set of
subsets $V = \{f_1,f_2,...f_m\}$. Wlog, we assume
in all cases
that each element is contained in at least two sets.
 With each subset in $V$ we
associate a value in the set $K = \{k_1,k_2,...k_m\}$.
With each element in $U$ we associate a value in the
set $P = \{v_1,v_2,...v_n\}$. We also define a set
of values $J = \{v_{ij}~|~1\leq i\leq n,~1\leq j\leq p\}$,
where $p$ is one greater than the number of attributes in some
R-component $Z$ of $m_2$.
The instance will also
contain a value $b$.

Relation $R$ ($S$) contains a set $S_{ij}^R$ ($S_{ij}^S$) of tuples for each
value $v_{ij}$ in $J$. Specifically, there is a tuple in
$S_{ij}^R$ and $S_{ij}^S$ for each set to which
$e_i$ belongs.
On attributes in $L$, all tuples in $S_{ij}^R$ and $S_{ij}^S$
take the value $v_{ij}$. There is a
tuple in $S_{ij}^R$ and a tuple in $S_{ij}^S$ for each value in $K$ corresponding to a set to
which $e_i$ belongs that has that value as the value
of all attributes in the R-component
of $m_1$ that contains an attribute in $E$.
On all other attributes, all tuples in all $S_{ij}^R$ and $S_{ij}^S$
take the value $b$.

Relation $S$ also contains a set $G_1$ of $m$
other tuples. For each value in $K$, there is a
tuple in $G_1$ that takes this value on all
attributes $A$ such that there
is an attribute $B\in E$ such that $B\approx A$
occurs in $m_2$. This tuple also takes this
value on all attributes of $S$ in $Z$. For all other attributes,
all tuples in $G_1$ take the value
$b$.

Relation $R$ also contains a set $G_2$ of $m$
other tuples. For each value in $K$, there is a
tuple in $G_1$ that takes this value on all
attributes in $E$ and all attributes of $R$ in
$Z$. Tuples in $G_2$ take the value $a$ on
all attributes in $L$. For all other attributes,
all tuples in $G_1$ take the value
$b$.

A resolved instance is obtained in two updates.
We first describe a sequence of updates that
will lead to an MRI, which we call our candidate update process.
It is easy to verify that the equivalence classes
of $T_{m_1}$ are the sets $S_{ij}^R\cup S_{ij}^S$.
In the first
update, the effect of applying $m_1$ is to update to a common value
all modifiable positions of attributes in $\nit{RHS}(m_1)$
for each equivalence class (c.f. Lemma \ref{lem:tc}).
For some minimum cover set $C$, we choose
as the update value for $S_{ij}^R\cup S_{ij}^S$ for all $j$
a value $k$ in $K$ that is associated with a set in $C$
containing $e_i$.

Before the
first update, there is one equivalence class of
$T_{m_2}$ for each value in $K$.
Let $E_k$ be the equivalence class for the value
$k\in K$.
$E_k$ contains all the
tuples in $R$ with $k$ as the value for the attributes
in $E$, as well as a tuple in $G_1$ with $k$ as
the value for the attributes in $Z$. We choose $k$ as the update value
for the modifiable positions of attributes
in $\nit{RHS}(m_2)$ for $E_k$.

After the first update, applying $m_1$ has
no effect, since none of the positions of attributes
in $\nit{RHS}(m_1)$ are modifiable. For each
update value that was chosen for the modifiable attributes
of $\nit{RHS}(m_1)$ in the first update there is
an equivalence class of $T_{m_2}$ that contains the union over all
sets $S_{ij}^R$ whose tuples' $\nit{RHS}(m_1)$ attributes were updated to that value
as well as the tuple of $G_1$ containing the
value. Given the choices of update values in the
previous update, it is easy to see that the
positions of attributes in $\nit{RHS}(m_2)$
that were modifiable before the first update are
modifiable after the first update. Thus, the
first update is ``overwritten" by the second.
We choose $b$ as the update value for the equivalence
classes of $T_{m_2}$ in the second update.

We now show that (i) our sequence of updates leads to
an MRI, and (ii) in an MRI, none of the positions of attributes of $S$ in
$Z$ for tuples in $G_1$ can have their values differ from the
original value, unless the value corresponds to
a cover set. (i) and (ii) together imply that a value of an attribute of $S$ in $Z$
for a tuple in $G_1$ is changed in some MRI iff the value corresponds to
a cover set.

Consider an arbitrary sequence of two updates.
When $m_1$ is applied to the instance during the first
update, the set of modifiable positions of attributes in
$\nit{RHS}(m_1)$ for each set $S_{ij}^R\cup S_{ij}^S$ of tuples
is updated to a common value. Our update sequence satisfies
the two conditions that (a) in the update resulting from
applying $m_1$, the update value chosen for all $S_{ij}^R\cup S_{ij}^S$
is the value in $K$ of a subset to which $e_i$ belongs and (b)
after the second update, all tuples in all
$S_{ij}^R\cup S_{ij}^S$ have the value $b$ for all attributes in
$\nit{RHS}(m_2)$. In an arbitrary update sequence, these conditions will generally be
satisfied only for some pairs $(i,j)$ of indices. Let $I$ be the
set of all pairs that satisfy (a) and (b). It is easy to verify that, in the resulting resolved instance,
for all $(i,j)$
not in $I$, the number of changes to positions of tuples in
$S_{ij}^R\cup S_{ij}^S$ is at least one greater than in our candidate update process.

First, we show that, for an MRI, $I$ must include all pairs $(i,j)$.
To prove this, we first show that, for any fixed value $i^*$, either all $(i^*,j)$
are in $I$ or none of them are. Suppose only some of the $(i^*,j)$
are in $I$. Suppose the update sequence is modified so that for all
$(i^*,j)$ not in $I$, the tuples in $S_{i^*j}^R\cup S_{i^*j}^S$ are
instead updated the same way as $S_{i^*j^*}^R\cup S_{i^*j^*}^S$
for some $(i^*,j^*)\in I$.
Then the number of changes to tuples in $S_{i^*j}^R\cup S_{i^*j}^S$
for each $(i^*,j)\notin I$ decreases by at least one, while the number
of changes to other tuples is unchanged.

Suppose that there exists an MRI $M$ such that there is an $(i^*,j^*)\notin I$.
By the preceding paragraph, $(i^*,j)\notin I$ for all $j$. Consider
a modification of the update sequence used to obtain $M$
that updates the tuples in $S_{i^*j}^R\cup S_{i^*j}^S$, $1\leq j\leq p$, according to
our candidate update process, while leaving all other updates the same.
This new update process will make at least $p$ fewer changes to the
tuples in $S_{i^*j}^R\cup S_{i^*j}^S$, $1\leq j\leq p$ (at least one fewer for
the tuples in each $S_{i^*j}^R\cup S_{i^*j}^S$). Furthermore, it can make at most
$p-1$ additional changes to positions in other tuples. This is because the only other tuples
that are updated as a result of having their values merged with those of the tuples in
$S_{i^*j}^R\cup S_{i^*j}^S$ are the tuples in $G_1$ and $G_2$ containing the value in $K$
that was used to update the tuples in $S_{i^*j}^R\cup S_{i^*j}^S$ according to $m_1$.
The values modified in these tuples are those of attributes in the R-component $Z$,
of which there are $p-1$.
Thus, the number of changes decreases as a result of changing the update process, contradicting
the statement that $M$ is an MRI.

For an MRI $M$, let $H$ be the set of update values used when applying $m_1$ to the tuples in
$S_{i^*j}^R\cup S_{i^*j}^S$. For each value in $H$, the positions of all attributes
in $Z$ are modified in $M$ for the tuple in $G_1$ that takes this
value on all attributes in $Z$. Since these are the only positions that are modified
by applying $m_2$, there are no more than $|H|\cdot (p-1)$ changes to the value positions of
attributes in $\nit{RHS}(m_2)$ in the second update used to produce $M$.
Therefore, $|H|$ must be as small as possible, implying
that $H$ corresponds to a minimum set cover. Furthermore, no
 other positions can be updated besides the ones updated during
the second update. This proves (i) and (ii).

Let $\mathcal{Q}$ be a query as in the statement
of the theorem. Let $k$ be the value in $K$ corresponding
to the candidate cover set in the CS instance.
We construct an assignment
to the free variables of $\mathcal{Q}$ as follows.
For some join-restricted free occurrence of the predicate $S$ ($R$),
assign to its variables the values of the tuple
in $G_1$ ($G_2$) whose value for the attributes in $Z$ is $k$.
For all other variables, assign the value $c_d$,
where $d$ is the domain of the associated attribute.
By construction, this assignment satisfies $\mathcal{Q}$
for all MRIs iff $k$ does not correspond to a cover set.

Let $\mathcal{Q}:= \ex\bar x\mathcal{Q}'$, with $\mathcal{Q}'$
a conjunction of atoms, be a query as in the statement
of the theorem. Let $k$ be the value in $K$ corresponding
to the candidate cover set in the CS instance.
We construct an assignment
to the free variables of $\mathcal{Q}$ as follows.
We construct an assignment
to the variables of $\mathcal{Q}'$ as follows.
For some join-restricted occurrence of the predicate $S$ ($R$),
assign to its variables the values of the tuple
in $G_1$ ($G_2$) whose value for the attributes in $Z$ is $k$.
For all other variables, assign the value $c_d$,
where $d$ is the domain of the associated attribute.
By construction, this assignment satisfies $\mathcal{Q}'$
for all MRIs if $k$ does not correspond to a cover set.
The converse is obvious.

Case (1)(b): This case uses the same set of values
as (1)(a). The instance is the same, except that the
tuples in $S_{ij}^R$ and $S_{ij}^S$ that took a certain value on
attributes in the R-component
of $m_1$ that contains an attribute in $E$ now take that value on
{\it all} such R-components. The update sequence that we specify
for obtaining an MRI is
also the same, but we add the requirement that
for a given equivalence class of $T_{m_1}$, the update
value must be the same for all R-components of $m_1$.

The difference between this case and (1)(a) is that different
update values can be chosen for different R-components of $m_1$
for the same equivalence class of $T_{m_1}$. It is easy to
verify that if different values are chosen, all tuples in the equivalence class
would be in singleton equivalence classes of $T_{m_2}$ after
the first update. Therefore, any changes made to positions of attributes in
$\nit{RHS}(m_2)$ for tuples in the equivalence class in the first update
cannot be undone in the second update.

Let $X$ denote the set of all $(i,j)$ such that there are two
R-components of $m_1$ that are updated to different values
for tuples in $S_{ij}^R\cup S_{ij}^S$. For $(i,j)\notin X$,
we use the same criteria as in part (1)(a) to classify
$(i,j)$ as being in $I$ or not.
For some $(i,j)\in X$, consider the update
values chosen for tuples in $S_{ij}^R\cup S_{ij}^S$
when $m_2$ is applied during the first update.
If any of the update values are not $b$, then, by the last sentence
of the preceding paragraph, at least one more change is made to
the tuples in $S_{ij}^R\cup S_{ij}^S$ than in our candidate update
process. In this case, we say $(i,j)$ is not in $I$. Otherwise,
$(i,j)$ is in $I$. The remainder of the proof is the same as in part (1)(a),
except that $H$ also contains, for each $(i,j)\in X\cap I$, all the values from
$K$ in tuples in $S_{ij}^R$.

Case (2): For simplicity of the presentation,
we will assume that there exists only one
attribute $A$ in $E$ not in $m_1$. If there is more than one such attribute, then all
tuples will take the same values on all such attributes as on $A$
in the instance produced by the reduction. Let $F$ be the min set cover instance
from case (1)(a), and define all sets of values
as before. We also have value $a$.
In addition, we define a set $Y$ of $2m^2np^2$ values,
which we denote by $y_{ij}$, $1\leq i\leq 2mnp^2$, $1\leq j\leq m$.
We also define a set $X$ of $2mnp^2$ values.

Relations $R$ and $S$ contain sets $S_{ij}^R$ and $S_{ij}^S$ for each $e_i$,
$1\leq i\leq n$,
as before. However, $S_{ij}^R$ and $S_{ij}^S$ now contain two
tuples for each set to which $e_i$ belongs.
On attributes
in $L$, tuples in each $S_{ij}^R$ and $S_{ij}^S$ take the same value as in
case (1)(a). Let $K' = \{k_1',k_2',...k_{|S_{ij}^R|/2}'\}$
and $K'' = \{k_1'',k_2'',...k_{|S_{ij}^R|/2}''\}$ be lists of
all the values in $K$ corresponding to sets to
which $e_i$ belongs such that $k_i' = k_{i\bmod{|S_{ij}^R|/2}+1}''$. For each value $k_i'\in K'$, there
are two tuples in $S_{ij}^R$ and two in $S_{ij}^S$ that take this value on all
attributes in all R-components of $m_1$ containing
an attribute of $E$. On the attribute $A$, one of
the two tuples in $S_{ij}^R$ takes the value $k_i'$ and the other
takes the value $k_i''$. (This ensures that the
tuples that take the value $k_i''$ will be in singleton equivalence
classes of $T_{m_2}$ before the first update.) On all other attributes,
all tuples in all $S_{ij}^R$ and $S_{ij}^S$
take the value $b$.

Relation $S$ ($R$) also contains a set $G_1$ ($G_2$) of $m$ tuples, which is the same
as the set $G_1$ ($G_2$) from case (1)(a).

Relation $R$ also contains a set $G_3$ of $2m^2np^2\cdot (2mnp^2+1)$ other tuples.
For each value $y_{ij}\in Y$, there is a set $Y_{ij}$ of $2mnp^2+1$ tuples that have
this value as the value of all attributes of $R$ in $L$. For each value
in the set $X$,
there is a tuple in $Y_{ij}$ that takes this value on attribute $A$.
On all other attributes in $E$, these tuples take the value $a$. On attributes in
$Z$, they take the value $k_j$ from the set $K$. On all other attributes
they take the value $b$.
There is also a tuple in $Y_{ij}$ that takes the value $k_j$
on all attributes in $E$ and on all attributes in $Z$. On all other attributes,
this tuple takes the value $b$.

Relation $S$ also contains a set $G_4$ of $2m^2np^2$ tuples. For each value in
$Y$, there is a tuple in $G_4$ that takes this value on all attributes of $S$
in $L$. On attributes in R-components of $m_1$ that contain an attribute in $E$,
all tuples in $G_4$ take the value $a$. On all other attributes, they take the
value $b$.

We now describe an update sequence that leads to an MRI, which
we call our candidate update process.
In this sequence, there are equivalence classes of $T_{m_1}$
that are the sets $S_{ij}^R\cup S_{ij}^S$, as in cases (1)(a) and (1)(b),
and we choose the update values for these equivalence classes
in the same way as in those cases.
There are also equivalence classes of $T_{m_1}$ that
involve
tuples in $G_3$ and $G_4$. Each of these consists of one of the $Y_{ij}$ sets
and the tuple in $G_4$ containing $y_{ij}$. We use $a$ as the
update value for these equivalence classes. This results
in all tuples in $G_3$ being in singleton equivalence classes
of $T_{m_2}$ after the first update.

Before the
first update, there is one equivalence class of
$T_{m_2}$ for each value in $K$.
Let $E_{k_j}$ be the equivalence class for the value
$k_j\in K$.
$E_{k_j}$ contains all the
tuples in $S_i^R$ and $G_3$ with $k_j$ as the value for all of the attributes
in $E$ (including $A$), as well as the tuple in $G_1$ with $k_j$ as
the value for attributes in $Z$. We choose $k_j$ as the update value
for the modifiable positions of attributes
in $\nit{RHS}(m_2)$ for $E_{k_j}$.

After the first update, the equivalence classes of
$T_{m_2}$ include the two tuples in $S_{ij}^R$ that
have the value to which their $\nit{RHS}(m_1)$ attributes
were updated in the first update as the value of
$A$. We choose $b$ as the update value for these
equivalence classes. Note that the fact that one
of the two tuples was in a singleton equivalence class
of $T_{m_2}$ before the first update guarantees that
all their positions that were modified by application of $m_2$ during the
first update are modifiable during the second update.

We claim that for an MRI, (i) the update values chosen in the first update for equivalence
classes of $T_{m_2}$ must be the same as in our candidate update process, and
(ii) the update values chosen for the equivalence classes of $T_{m_1}$ containing
tuples of $G_3$ must be the same as in our candidate update process.
Statement (ii) follows from the fact that, if any value other than
$a$ is chosen as the update value for such an equivalence class, it
would result in at least $2mnp^2+1$ more changes to the positions of
attributes in $\nit{RHS}(m_1)$ than in our update sequence. Since
our candidate update process makes no more than $2mnp^2$ changes to the positions of
attributes in $\nit{RHS}(m_2)$, no such alternative update sequence
could produce an MRI. Similarly, (i) follows from the fact that,
if for any $E_{k_j}$, any value other than $k_j$ is chosen as the
update value, there would be at least $2mnp^2$ changes to positions
of attributes in $\nit{RHS}(m_2)$ for tuples in $G_3\cap E_{k_j}$ during the first update. If that is the case, then
some of these positions must be restored to their original values in the
second update. However, this would require some of the tuples in $G_3\cap E_{k_j}$
to be in non-singleton equivalence classes of $T_{m_2}$ after the first update,
which by (ii) is not possible.

Consider an arbitrary sequence of two updates.
When $m_1$ is applied to the instance during the first
update, the set of modifiable positions of attributes in
$\nit{RHS}(m_1)$ for each set $S_{ij}^R\cup S_{ij}^S$ of tuples
is updated to a common value. Our candidate update process satisfies
the three conditions that (a) in the update resulting from
applying $m_1$, the update value chosen for each set $S_{ij}^R\cup S_{ij}^S$
is the same for all R-components, (b) this update value
is the value in $K$ of a subset to which $e_i$ belongs and (c)
in the second update, the update value chosen for modifiable positions in tuples in each set
$S_{ij}^R$ is $b$. In an arbitrary update sequence, these conditions will generally be
satisfied only for some pairs $(i,j)$ of indices. Clearly, for pairs of indices
not satisfying (b), there will be at least one more change to the values of
tuples in $S_{ij}^R$ than in our candidate update process. Given (i)
above, this is also true for pairs of indices not satisfying (c).
For pairs of indices not satisfying (a), all tuples in $S_{ij}^R$
are in singleton equivalence classes of $T_{m_2}$, and therefore
(c) cannot be satisfied. Therefore, failing to satisfy any of (a),
(b), and (c) results in at least one more change to the values of
tuples in $S_{ij}^R$ than in our candidate update process. Let
$I$ be the set of all pairs that satisfy (a), (b), and (c). We now
use exactly the same argument involving the set $I$ as in part (1)(a)
to prove the result.

Case (3): Let $F$ be the CS instance
from case (1)(a), and define sets of values $K$
and $P$ as before. Let $E'$ be an ES containing attributes of $m_1$. Since
the MDs are interacting, there must be at least
one such ES, and by assumption, it
must contain an attribute of $\nit{LHS}(m_1)$. Let $C_1$
denote some R-component of $m_1$ that contains
an attribute of $E'$, and let $p$ denote the number
of attributes in $C_1$. Let $C_2$ denote some R-component
of $m_2$. Let $q_R$ and $q_S$ be
the number of attributes of $R$ and $S$ in $C_2$, respectively.
Let $d_i$ be the number of elements in the set $f_i$.
We define a set $W_j$ of values of size $4q_Sp^2$
for each $j$ such that $e_j\in f_i$. We also define
sets $Y_{ij}$ and $Z_{ij}$ of $p$ values
each and $4q_Sq_R$ values each, respectively, for all pairs
of indices $i,j$ such that $e_j\in f_i$.
We also define set $X$ containing $nq_S$ values,
 and values $a$ and $b$.

Relation $R$ ($S$) contains a set $S_i^R$ ($S_i^S$) for each set $f_i$,
$1\leq i\leq m$, in $V$. For each element $e_j$ in $f_i$,
$S_i^R$ ($S_i^S$) contains a set $S_{ij}^R$ ($S_{ij}^S$) of $4q_Sp+4q_Sq_R$ tuples. On all attributes
of $L$, all tuples in $S_i^R$ and $S_i^S$ take the value $k_i$
in $K$ corresponding to $f_i$. For each $S_{ij}^R$ and $S_{ij}^S$,
there is a set of $4q_Sp$ tuples $S_{ij}^{R*}$ in $S_{ij}^R$ and a set
of $4q_Sp$ tuples $S_{ij}^{S*}$ in $S_{ij}^S$.
For each $i$, for all $j$ such that $e_j\in f_i$ except one, each value
in $W_j$ occurs once as the value of an attribute of either $R$ or $S$
in $C_1$ for a tuple in $S_{ij}^{R*}$ (or $S_{ij}^{S*}$). For the
remaining $j$, all but two of the values in $W_j$ occur
once as the value of an attribute of either $R$ or $S$
in $C_1$ for a tuple in $S_{ij}^{R*}$ (or $S_{ij}^{S*}$).
This leaves the values of two positions of tuples in
$S_{ij}^{R*}$ and $S_{ij}^{S*}$ for attributes in $C_1$ undefined,
and two values in the set $W_j$ unassigned.
These positions take the same value, which is one
of the unassigned values from $W_j$. We call this
value $s_i$. All other tuples
in $S_{ij}^R$ and $S_{ij}^S$ take the value $a$ on all attributes
in $C_1$.

For each value
in $Y_{ij}$, there are $4q_S$ tuples in $S_{ij}^{R*}$ that
take the value on all attributes in $C_2$.
For each value in $Z_{ij}$, there is a tuple in
$S_{ij}^R$ not in $S_{ij}^{R*}$ that
takes the value on all attributes in $C_2$.
On all attributes of $E$, each tuple in $S_{ij}^{R*}$,
$1\leq i\leq m$, takes the value $v_j$ in $P$ that
is associated with $e_j$, and all other tuples in $S_{ij}^R$
take the value $b$. On all other attributes,
all tuples in $S_i^R$ and $S_i^S$ take the value $a$.

Relation $S$ also contains a set of $n$ tuples $G_1$.
For each value in
$X$, there is a tuple in $G_1$ that takes the
value on an attribute of $C_2$. For each value
in $P$, there is a tuple in $G_1$ that takes this value
on all attributes
$B$ such that there is an attribute $A$ of $R$ in
$E$ such that $A\approx B$ is a conjunct of $m_2$. On all other
attributes, tuples in $G_1$ take the value $a$.

A resolved instance is obtained in two updates.
Before the first update, the equivalence classes of $T_{m_2}$
are all singletons. The equivalence
classes of $T_{m_1}$ are the sets $S_i^R\cup S_i^S$, $1\leq i\leq n$. The effect of applying $m_1$
is to change all positions of all attributes
in $C_1$ for tuples in $S_i^R\cup S_i^S$ to a common value. It
is easy to verify that if the update value is not $a$, then
all tuples in $S_i^R$ will be in singleton equivalence
classes of $T_{m_2}$ after the update. Thus, the equivalence
classes of $T_{m_2}$ after the update are
$\bigcup_{i\in I} S_{ij}^R\bigcup x_j$, $1\leq j\leq n$, where
$I \equiv\{i~|~a\hbox{ was chosen as the update value for }
S_i^R\cup S_i^S\}$ and $x_j$ is the tuple in $G_1$ containing the
value $v_j$.
If the update value $a$ is chosen for set $S_i^R\cup S_i^S$ for some $i$,
 we say that $S_i^R$ is
{\em unblocked}. Otherwise, it is {\em blocked}.

For a given $i$, we will consider the
number of changes resulting from different choices
of update values for tuples in $S_i^R\cup S_i^S$. These
changes include all changes that are affected by the
choice of update value for tuples in $S_i^R\cup S_i^S$.

Consider a blocked $S_i^R$.
In the first update, the minimum number of changes
to positions of attributes in $\nit{RHS}(m_1)$
for tuples in $S_i^R\cup S_i^S$ is $4q_Spd_i(p+q_R) - 2$,
where $d_i$ is the number of elements in $f_i$.
All tuples in $S_i^R$ are in singleton equivalence classes of
$T_{m_2}$ before and after the first update.
 Therefore, the number of changes resulting from this choice
of update value is $4q_Spd_i(p+q_R) - 2$.

For an unblocked $S_i^R$, the minimum number of changes
to values for attributes in $\nit{RHS}(m_1)$ for tuples in
$S_i^R\cup S_i^S$ is $4q_Sp^2d_i$.
A set $S_{ij}^R$ is {\em good} if, in the second update, all positions in the set
of positions of attributes in $C_2$ for tuples in $S_{ij}^R$ are modified
to the value of a position in the set. The set $S_i^R$ is {\em good}
if it contains a good $S_{ij}^R$. Sets $S_{ij}^R$ and
$S_i^R$ that are not good are {\em bad}. The total number of
changes to positions of attributes of $\nit{RHS}(m_2)$
for tuples in
a bad unblocked $S_i^R$ is $4q_Sq_Rpd_i$, and
for a good unblocked $S_i^R$ it is $4q_Sq_Rpd_i - 4q_Sq_Rg_i$,
where $g_i$ is the number of good $S_{ij}^R$ in $S_i^R$.
Thus, the total number of changes for the bad
case is $4pq_Sd_i(p+q_R)$ and for the good case it is
$4pq_Sd_i(p+q_R) - 4q_Sq_Rg_i$.

The number of changes to tuples in a bad unblocked $S_i^R$ is
larger than that in tuples in a blocked $S_i^R$.
Since tuples in a blocked $S_i^R$ are in singleton
equivalence classes of $T_{m_2}$ both before and after
the first update, choosing a $S_i^R$ to be blocked also minimizes the
number of changes to tuples not in $S_i^R$.
Therefore, in
an MRI, all unblocked $S_i^R$ are good.
 Let $U$ ($B$) be the set of $i$ for which $S_i^R$ is unblocked (blocked).
For an MRI, the total number of changes for all $S_i^R$ is
\bea
\sum_{i\in U}[4pq_Sd_i(p+q_R)-4q_Sg_iq_R] + \sum_{i\in B}[4pq_Sd_i(p+q_R)-\nn\\
2]\nn
\eea
plus the number of changes to tuples in $G_1$. To compute the
latter, we note that values in a tuple in $G_1$ can change iff
the tuple contains $v_j$, where $e_j\in f_i$ for some $i\in G$.
In this case, there must be some $i^*\in U$ such that $S_{i^*j}$
is good. Indeed, if this were not the case, then the tuple
in $G_1$ containing $v_j$ would always be in a singleton equivalence class
of $T_{m_2}$. Therefore
the number of tuples in $G_1$ that change is $\sum_{i\in G}g_i$,
and the total number of changes is
\bea\label{eq:change}
&&\sum_{i\in U}[4pq_Sd_i(p+q_R)-4q_Sg_iq_R+q_Sg_i] + \nn\\
&&\sum_{i\in B}[4pq_Sd_i(p+q_R)-2]\nn\\
&&= 4pq_S(p+q_R)\sum_{1\leq i\leq m}d_i - (4q_Sq_R-q_S)\cdot  \\
&&\sum_{i\in U}g_i-2|B|\nn
\eea

The first term in (\ref{eq:change}) depends only on the database
instance and not on the choice of update values. Therefore, the
number of changes is minimized by choosing the update values so
as to maximize the magnitude of the last two terms.

The sum over $g_i$ in the second term in (\ref{eq:change})
is bounded above by $n$. This can be shown as follows. After the first
update, there is one
equivalence class for each value of $j$, containing the
set of all $S_{ij}^R$ such that $S_i^R$ is unblocked. Furthermore,
the sets of values of modifiable positions
of attributes in $RHS(m_2)$ for tuples in a given $S_{ij}^R$ do not
overlap with those of any other $S_{ij}^R$. Therefore,
at most one $S_{ij}^R$ can be good for any value of $j$.

If the sum over $g_i$ equals $n$, then the set of subsets corresponding to
the set of $i$ for which $S_i^R$ is good is a set cover.
If it is a min set cover, then $|B|$ is maximized for this
value of the sum.

We claim that the magnitude of the last two terms in (\ref{eq:change})
is maximized by choosing the set of good $S_i^R$ so
that $\{e_i~|~S_i^R\hbox{ is good}\}$ is a min set cover,
from which it follows that this choice is required for the
resolved instance to be an MRI. Suppose for a contradiction that
there is an MRI $M$ for which the sum over $g_i$ in (\ref{eq:change})
is $n-c$ for some $1\leq c\leq n$. This implies that, for $M$, there
is a set $J$ of $c$ values of $j$ such that there is no $i$ such that
$S_{ij}^R$ is good. Consequently, for $j^*\in J$, for any $i$ such that
$S_{ij^*}^R$ exists, $S_i^R$ must be blocked. This is because if there
were an unblocked $S_i^R$, then the second update could be changed so
that $S_{ij^*}$ is good, reducing the number of changes.

We modify the update sequence used to obtain $M$ in the following way.
For each $j\in J$, choose an $i$ such that $S_{ij}^R$ exists. For each
such $i$, change the first update so that $S_i^R$ is unblocked, and
change the second update so that $S_{ij}^R$ is good. This will
increase the magnitude of the second term in (\ref{eq:change})
by $(4q_Sq_R-q_S)c\geq 3c$ and decrease the magnitude of the third term by at most
$2c$. Therefore, the number of changes decreases as a result of this
modification to the update sequence, contradicting the assumption
that $M$ is an MRI.

The value $s_i$ is the value of an attribute in $C_1$ for
a tuple in $R$ or $S$ iff $f_i$ is not a cover set. The remainder
of the proof is similar to the last paragraph of the proof
for case (1)(a).
}

\vspace{2mm}
\defproof{Theorem \ref{thm:dichotomy}}{
We assume that an attribute of both $R$ and $S$
in $\nit{RHS}(m_1)$ occurs in $\nit{LHS}(m_2)$.
The other cases are similar.
For each L-component of $m_1$,
there is an attribute of $R$ and an attribute of $S$
from that L-component
in $\nit{LHS}(m_2)$. Let $t_1\in R$ be a
tuple not in a singleton equivalence class of $T_{m_1}$.
Suppose there exist two conjuncts in $\nit{LHS}(m_1)$ of the
form $A\approx B$ and $C\approx B$. Then it must hold that
there exists $t_2\in S$ such that
$t_1[A]\approx t_2[B]$ and $t_1[C]\approx t_2[B]$ and by
transitivity, $t_1[A]\approx t_1[C]$. More generally, it follows
from induction that $t_1[A]\approx t_1[E]$ for
any pair of attributes $A$ and $E$ of $R$ in the same L-component of $m_1$.

We now prove that for any pair of tuples $t_1,t_2\in R$
satisfying $T_{m_2}(t_1,t_2)$ such that each of
$t_1$ and $t_2$ is in a non-singleton equivalence class
of $T_{m_1}$, for any instance $D$ it
holds that $T_{m_1}(t_1,t_2)$. By symmetry, the same result
holds with $R$ replaced with $S$.
Suppose for a contradiction that $T_{m_2}(t_1,t_2)$ but
$\neg T_{m_1}(t_1,t_2)$ in $D$.
 Then it must be true that
$t_1[\bar A]\not\approx t_2[\bar A]$, since, by assumption,
there exists a $t_3\in S$ such that
$t_1[\bar A]\approx t_3[\bar B]$, which together with $t_1[\bar A]\approx t_2[\bar A]$
would imply $T_{m_1}(t_1,t_2)$. Therefore, there must be an
attribute $A'\in \bar A$ such that $t_1[A']\not\approx t_2[A']$,
and by the previous paragraph and transitivity, $t_1[A'']\not\approx t_2[A'']$ for
all $A''$ in the same L-component of $m_1$ as $A'$. By
transitivity of $\approx_2$, this implies $\neg T_{m_2}(t_1,t_2)$,
a contradiction.

A resolved instance is obtained in two updates.
Let $T_{m_2}^0$ and $T_{m_2}^1$ denote $T_{m_2}$
before and after the first update, respectively.
The first update involves setting the attributes in $\nit{RHS}(m_1)$
to a common value for each non-singleton equivalence class of $T_{m_1}$.
The relation $T_{m_2}^1$ will depend on these common
values, because of accidental
similarities. However, because of the property
proved in the previous paragraph, this dependence
is restricted. Specifically, for each equivalence
class $E$ of $T_{m_2}^1$, there is at most
one non-singleton equivalence class $E_1$ of $T_{m_1}$
such that $E$ contains tuples of $E_1\bc R$ and
at most one non-singleton equivalence class $E_2$ of $T_{m_1}$
such that $E$ contains tuples of $E_1\bc S$.
 A given choice of update values
for the first update will result in a set of sets of tuples
from non-singleton equivalence classes of $T_{m_1}$ (ns tuples)
that are equivalent under $T_{m_2}^1$.
Let $K$ be the set of all such sets of
ESs. Clearly, $|K|\in O(n^2)$,
where $n$ is the size of the instance.

Generally, when the instance is updated according to
$m_1$, there will be more than one set of choices of
update values that will lead to the ns tuples being
partitioned according to a given $k\in K$. This is
because an equivalence class of $T_{m_2}^1$ will also
contain tuples in singleton equivalence classes of $T_{m_1}$
(s tuples), and the set of such tuples contained in
the equivalence class will depend on the update values
chosen for the modifiable attribute values in the ns tuples
in the equivalence class.
For a set $E\in k$, let $E'$ denote
the union over all sets of update values for $E$ of the equivalence classes
of $T_{m_2}^1$ that contain $E$ that result from choosing that set of
update values. By transitivity and
the result of the second paragraph, these $E'$ cannot overlap
for different $E\in k$.
Therefore, minimization of the change produced by the two
updates can be accomplished by minimizing the change for
each $E'$ separately. Specifically, for each equivalence class
$E$, consider the possible sets of update values for the attributes in
$\nit{RHS}(m_1)$ for tuples in $E$. Call two such sets of values equivalent if they
result in the same equivalence class $E_1$ of $T_{m_2}^1$.
Clearly, there are at most $O(n^c)$ such sets of ESs of values,
where $c$ is the number of R-components of $m_1$.
Let $V$ be a set consisting of one set of values $v$ from each set of sets of equivalent values.
For each set of values $v\in V$, the minimum number of changes produced
by that choice of value can be determined as follows.
The second application of $m_1$ and $m_2$ updates to a common
value each element in a set $S_2$ of sets of value positions that can be
determined using
lemma \ref{lem:tc}. The update values that result in
minimal change are easy to determine. Let $S_1$ denote the corresponding set of sets of value positions
for the first update. Since the second update ``overwrites" the first,
the net effect of the first update is to change to a common value the value
 positions in
each set in $\{S_i~|~S_i = S\backslash \bigcup_{S'\in S_2} S',~S\in S_1\}$.
It is straightforward to determine the update values that yield
minimal change for each of these sets.
This yields the minimum number of changes for this choice of $v$. Choosing
$v$ for each $E$ so as to minimize the number of changes allows the
minimum number of changes for resolved instances in which
the ns tuples are partitioned according to $k$ to be determined
in $O(n^c)$ time. Repeating
this process for all other $k\in K$ allows the determination
of the update values that yield an MRI in $O(n^{c+2})$ time.
Since the values to which each value in the instance can change
in an MRI can be determined in polynomial time, the result
follows.}

\vspace{2mm}
\defproof{Theorem \ref{thm:main}}{
For simplicity, we prove the theorem for the
special case in which $M$ is defined on a single relation $R$
and both attributes in each conjunct
are the same. The same argument can be used for arbitrary sets
of pair-preserving MDs by adding the additional restriction that the set $I$ defined
below contains only instances for which the set of values
taken by pairs of attributes occurring in the same conjunct
are the same.

The proof is by reduction from the
resolved answer problem for a set of MDs that is hard
by Theorem \ref{thm:combine}.
Specifically, we will construct a set $I$ of database instances.
We then give a polynomial time reduction from
(a) the resolved answer problem for a specific pair of MDs to (b) the
current problem, where for both (a) and (b), the input to the
problem is restricted to having instances in $I$. We will show that
(a) remains intractable when instances are restricted to $I$. Since
(b) restricted to $I$ can obviously be reduced to the current problem
in polynomial time, this proves the theorem.

We define a set $S_1$ of attributes recursively according to Definition
\ref{def:inclusion}. An attribute $A$ is in $S_1$ if (a)
$A\in \nit{LHS}(m)$ for some $m$ such that $C\in \nit{RHS}(m)$,
(b) $A\nin \nit{LHS}(m_1)\bigcup \nit{LHS}(m_2)$
and (c) $A$ is non-inclusive wrt $\{m_1,m_2\}$, or if $A$ satisfies
(a), (b), and (c) with $C$ replaced by an attribute in $S_1$.
For all attributes $A\in S_1$, all values in the $A$ column
for instances in $I$ are dissimilar to each other.

The set $S_2$ of attributes is defined similarly. An attribute
$A$ is in $S_2$ if (a)
$A\in \nit{LHS}(m)$ for some $m$ such that $B\in \nit{RHS}(m)$,
 (b) $A\nin \nit{LHS}(m_2)$ (c) $A$ is non-inclusive wrt $\{m_2\}$, and (d) $A\nin S_1$,
 or $A$ satisfies
(a), (b), (c), and (d) with $B$ replaced by an attribute in $S_2$.
The second requirement for an instance to be in $I$ is that,
for any pair of tuples in the instance, the tuples are either
equal on all attributes in $S_2$ or dissimilar on all attributes
in $S_2$.

For all attributes not in $S_1$ or $S_2$ besides $B$ and $C$, all tuples in instances
in $I$ have the same value for the attribute.

Consider the set $M'$ of MDs
\bea
m_1':~R[E]\approx R[E]\ra R[B]\doteq R[B]\nn\\
m_2':~R[B]\approx R[B]\ra R[C]\doteq R[C]\nn
\eea
where $E\in \nit{LHS}(m_1)\bigcap S_2$ (there must be
such an $E$ by assumptions (a) and (b) of the theorem).
By Theorem \ref{thm:combine},
$M'$ is hard.
We claim that (1) $RA_{\mathcal{Q},M'}$ for a changeable attribute query
$\mathcal{Q}$ remains intractable when
input instances are restricted
to $I$, and (2) $RA_{\mathcal{Q},M'}$ for any $\mathcal{Q}$ reduces in
polynomial time to $RA_{Q,M}$ when input instances are restricted
to $I$. These two claims imply the theorem.

Claim (1) is true because the reduction in the proof of Theorem \ref{thm:combine}
can be made to always produce an instance in $I$
by making a specific choice of the values in the
instance that were allowed to be arbitrary in that proof. Specifically,
since $R[B]$ and $R[C]$ are not in $S_1\bigcup S_2$, the values
that tuples in instances in $I$ can take on attributes
$R[E]$, $R[B]$, and $R[C]$ are unrestricted. Given the
values that tuples in an instance in $I$ take on
$R[E]$, $R[B]$, and $R[C]$, the values that the
tuples can take on attributes not in $m_1$ and $m_2$
are restricted. However, in the proof of Theorem \ref{thm:combine},
the values for these attributes in the instance produced
by the reduction were (mostly) left unspecified, and
it is easily verified that they can always be chosen so that
this instance is in $I$.

To prove claim (2), we show that the set of all updates that
can be made under $M'$ is the same as that under $M$, for
any instance in $I$. Thus, the reduction is simply the identity
transformation.

First, we show that, for any MD $m$ other than $m_1$ and $m_2$,
applying $m$ has no effect. Such MDs can therefore be ignored
when updating the instance.
If $\nit{RHS}(m)$ consists of an attribute not in $S_1\bigcup S_2$
and is not $B$ or $C$, then applying $m$ cannot change the values
of the attribute, because these values are already the same.
If $\nit{RHS}(m)$ is an attribute of $S_1$ ($S_2$), then by
definition of these sets, $\nit{LHS}(m)$ contains an attribute
of $S_1$ ($S_2$). Therefore, any pair of tuples satisfying the similarity
condition of $m$ must already have equal values for the attribute
in $\nit{RHS}(m)$, and applying $m$ has no effect. If $C$ is
the attribute in $\nit{RHS}(m)$, then there must be an attribute
of $S_1$ in $\nit{LHS}(m)$. Since all values for this attribute
are mutually dissimilar and are never updated,
no pair of tuples satisfies the similarity condition of
$m$, so applying $m$ has no effect. Lastly, if $B$ is
the attribute in $\nit{RHS}(m)$, we claim that
updates resulting from $m$ are subsumed by those resulting
from $m_1$. Indeed, by definition of $S_1$, there are no
attributes of $S_1$ in $\nit{LHS}(m_1)$, and by the acyclic
property, neither $B$ nor $C$ are in $\nit{LHS}(m_1)$.
Given this and the fact that there is an attribute of
$S_2$ in $\nit{LHS}(m)$ (by definition of $S_2$),
it is easy to verify that if a pair of tuples satisfies the
similarity condition of $m$, it must satisfy the similarity
condition of $m_1$.

We now show that the effect of applying $\{m_1,m_2\}$ to
an instance in $I$ is the same as that of applying $M'$ to
the instance, thus proving the theorem. All attributes
in $\nit{LHS}(m_1)$ are either in $S_2$ or have the
same value for all tuples. Thus, all tuples satisfying the
conjunct $R[E]\approx R[E]$ also satisfy all other conjuncts
to the left of the arrow in $m_1$. By definition, $\nit{LHS}(m_2)$
contains no attributes of $S_1$ and $S_2$, and by
the acyclic property, it does not contain $C$. Therefore,
all attributes besides $B$ in $\nit{LHS}(m_2)$ have the same
value for all tuples. This implies that all pairs of tuples
satisfying $R[B]\approx R[B]$ satisfy the similarity condition
of $m_2$.}

\defproof{Proposition \ref{prop:nontransitive}}{
We take finite strings of bits as the domain of all attributes.
We number each bit within a string consecutively from left to right
starting at 1.
Two strings are similar if they both have a 1 bit with the
same number. Otherwise, they are dissimilar. For example, 011 and
010 are similar, but 010 and 100 are dissimilar.
It is readily verified that this satisfies the properties of
a similarity operator.

As in the proof of Theorem \ref{thm:combine}, we reduce
the complement of CS to the resolved answer
problem for the given MDs. Let $F$ be an instance
of CS as in Case (1)(a) of the proof of Theorem \ref{thm:combine},
and define $U$, $V$, $K$, and $P$ as before. We take $v_i$
to be a string with $n+1$ bits, all of which are 0 except
the $i^{th}$ bit. We take $k_i$ to be a string with $m$
bits, with all bits 0 except the $i^{th}$ bit. The instance
will also contain strings $a$, $b$, and $c$ of length $n+1$.
String $a$ is all zeros, $b$ is all zeros except the
$(n+1)^{th}$ bit, and $c$ is all ones.

There are $n$ sets of tuples $S_i$, $1\leq i\leq n$,
which contain $|f_i|$ tuples of $R$ and $|f_i|$ tuples of $S$.
On attributes $R[A]$ and $S[B]$, the tuples in $S_i$
take the value $v_i$. On attributes $R[I]$ and $S[J]$,
all tuples in all $S_i$ take the value $c$. On attributes
$R[G]$ and $S[H]$, all tuples in all $S_i$ take the
value $a$. In each $S_i$, for each set $f_i$ to which $e_i$
belongs, there is one tuple in $R$ and one tuple in $S$
that has $k_i$ as the value of $R[E]$ (or $S[F]$).

There is also a set $G_1$ of $m$ tuples in $S$. On
$S[B]$, all tuples in $G_1$ take the value $b$.
On $S[J]$, all tuples in $G_1$ take the value $a$.
For each value in $K$, there is a tuple in
$G_1$ that takes the value on $S[F]$ and $S[H]$.

The result is now proved analogously to part (1)(a)
of the proof of Theorem \ref{thm:combine}.}

\end{document}